\DocumentMetadata{}

\documentclass[sigconf,screen,authorversion]{acmart}
\AtBeginDocument{%
  \providecommand\BibTeX{{%
    \normalfont B\kern-0.5em{\scshape i\kern-0.25em b}\kern-0.8em\TeX}}}

\usepackage{soul}
\usepackage{svg}
\usepackage{CJKutf8}
\usepackage{booktabs}
\usepackage{multirow}
\usepackage{siunitx}
\usepackage{makecell}
\usepackage{subcaption}

\colorlet{usercolorname}{yellow!0}
\sethlcolor{usercolorname}

\copyrightyear{2025}
\acmYear{2025}
\setcopyright{cc}
\setcctype{by-nc-sa}
\acmConference[CHI '25]{CHI Conference on Human Factors in Computing Systems}{April 26-May 1, 2025}{Yokohama, Japan}
\acmBooktitle{CHI Conference on Human Factors in Computing Systems (CHI '25), April 26-May 1, 2025, Yokohama, Japan}\acmDOI{10.1145/3706598.3713323}
\acmISBN{979-8-4007-1394-1/25/04}

\begin{document}
\begin{CJK}{UTF8}{ipxm}

\title[Inter(sectional) Alia(s)]{Inter(sectional) Alia(s): Ambiguity in Voice Agent Identity via Intersectional Japanese Self-Referents}

\author{Takao Fujii}
\email{fujii.t.av@m.titech.ac.jp}
\orcid{0009-0004-5059-3323}
\affiliation{%
  \institution{Institute of Science Tokyo}
  \city{Tokyo}
  \country{Japan}
}

\author{Katie Seaborn}
\email{seaborn.k.aa@m.titech.ac.jp}
\orcid{0000-0002-7812-9096}
\affiliation{%
  \institution{Institute of Science Tokyo}
  \city{Tokyo}
  \country{Japan}
}

\author{Madeleine Steeds}
\orcid{0000-0003-3767-292X}
\email{madeleine.steeds@ucd.ie}
\affiliation{%
  \institution{University College Dublin}
  \city{Dublin}
  \country{Ireland}
}

\author{Jun Kato}
\email{jun.kato@aist.go.jp}
\orcid{0000-0003-4832-8024}
\affiliation{%
  \institution{AIST}
  \city{Tokyo}
  \country{Japan}
}

\renewcommand{\shortauthors}{Fujii et al.}

\begin{abstract}
Conversational agents that mimic people have raised questions about the ethics of anthropomorphizing machines with human social identity cues. Critics have also questioned assumptions of identity neutrality in humanlike agents. Recent work has revealed that intersectional Japanese pronouns can elicit complex and sometimes evasive impressions of agent identity. Yet, the role of other ``neutral'' non-pronominal self-referents (NPSR) and voice as a socially expressive medium remains unexplored. In a crowdsourcing study, Japanese participants ($N=204$) evaluated three ChatGPT voices (Juniper, Breeze, and Ember) using seven self-referents. We found strong evidence of voice gendering alongside the potential of intersectional self-referents to evade gendering, i.e., ambiguity through neutrality and elusiveness. Notably, perceptions of age and formality intersected with gendering as per sociolinguistic theories, especially ぼく (boku) and わたくし (watakushi). This work provides a nuanced take on agent identity perceptions and champions intersectional and culturally-sensitive work on voice agents.
\end{abstract}





\maketitle



\section{Introduction}
ChatGPT, with its underlying large language model (LLM) trained on a massive amount of human-generated data, is now widely used globally. ChatGPT can converse naturally in many languages and in a seemingly humanlike way~\cite{NassSundar2000Source, marchandot2023chatgpt}.
Yet, this raises ethical concerns. 
The data used to train LLMs can carry biases related to gender, age, ethnicity/race, class, and other social identities~\cite{schlesinger2017intersectional, ciston2019intersectional}.
Developers may unintentionally bias the algorithm and data sets, leading to limited and stereotyped results~\cite{bolukbasi2016man,buolamwini2018gender}. Users may perceive human-derived cues in line with stereotypes~\cite{eyssel2012s, nag2020gender,perugia2023models, craiut2022technology}. 
This is important because identity is central to the ``human'' side of human-computer interaction (HCI). While machines do not have social identities, they can be treated as 
``social agents''~\cite{nass1997machines,NassSundar2000Source} and attributed human markers like gender~\cite{purington2017alexa}\footnote{Recognizing that AI-attributed ``identities'' are a matter of human perception, we use such phrasing as ``age perceptions'' and ``agedness.''}. 
Even so, certain limitations have prevailed in the work so far. Most work has aimed at uncovering biases in the system or the algorithms~\cite{buolamwini2018gender, karizat2021algorithmic}. Furthermore, an \emph{intersectional} approach that explains how power operates through multiple social identities~\cite{collins2022black,crenshaw2013mapping}, remains underexplored~\cite{ciston2019intersectional, schlesinger2017intersectional, Jarrell2021usinginter}, however crucial for grasping AI identity. 
Research examining social identity perceptions about such agents from an intersectional perspective is still nascent.

The social medium of \emph{voice} can influence identity perceptions~\cite{seaborn_can_2023,Seaborn2021Voice,unesco2019,McGinn2019voice}. Voice is multidimensional, involving vocalics~\cite{Seaborn2021Voice,seaborn_can_2023}, vocal prosody~\cite{Druga2017, Crumpton2016prosody}, non-semantic vocalizations~\cite{Yilmazyildiz2016semantic}, emotions and feelings~\cite{Seaborn2021Voice}, and social identity cues (e.g., gender, dialects, accents)~\cite{Swati2015emotion}.
Voice-based agents 
can elicit simple and complex \emph{personas}~\cite{Seaborn2021Voice, nass1997machines, LeeKimLee2019}, including gender-based ones~\cite{nag2020gender}. With widely-used agents like ChatGPT 
now offering a voice mode, research on human--agent communication~\cite{Jasper2019taxonomy,Julia2019smart} and the voice dimension~\cite{Seaborn2021Voice} may be  paramount. 
Notably, a 2019 UNESCO report~\cite{unesco2019} highlighted concerns about the negative implications of gendered voice-based agents, 
the majority of 
which are designed to be ``female exclusively or female by default''~\cite[p. 107]{unesco2019}. Indeed, a range of work \cite{Jungyong2022stereotypes, tay2014stereotypes, craiut2022technology} has shown that evaluations of an agent's competence are influenced by stereotypes of the agent's perceived gender. Still, other work has offered contrasts, with \citet{Thellman2018speeches} showing that speech persuasiveness was not influenced by robot voice or appearance  gendering, and \citet{Pfeuffer2019MrandMrs} showing that conversational agents with feminine gender cues ware perceived as more competent than their masculine-cued counterparts.
Moreover, how speech content and voice intersect, especially when one or the other is ambiguous, remains unclear. For instance, the Japanese ChatGPT ``Breeze'' voice has been labelled feminine\footnote{https://www.goatman.co.jp/media/chatgpt-voice-howtouse/} \emph{and} masculine\footnote{\url{https://note.com/it_navi/n/n70b9de7222d1}}, indicating its gender neutral or ambiguous potential.
At present, how gender in the conversationally fluent ChatGPT voices are perceived 
has yet to be researched.


ChatGPT also uses \emph{linguistic} social identity cues, notably \emph{self-referents} like \emph{first-person pronouns}~\cite{balmer2023sociological} like ``I'' in English. However, this raises ethical concerns. 
Pronouns are particularly easy to manipulate in machine-generated text~\cite{cho2019measuring, seaborn2023transc, SeabornFrank2022Pepper, seaborn2023imlost, sun2019mitigating, fujii2024silver} and can be biased by the data or training methods~\cite{cho2019measuring, seaborn2023imlost, sun2019mitigating}. 
Agents can use pronouns to elicit simple and complex personas
~\cite{nakamura2022feminist,fujii2024silver}.
Self-referents can enact diversity but they can also reinforce prejudices~\cite{fujii2024silver, nakamura2014gender, maree2007language} and stereotypes
~\cite{nakamura2020formation, hirano1994heterosexism, maree2007language}. 
In many languages, self-referents are pronouns linked to gender, e.g., 
English third-person pronouns like ``she'' and ``he''
~\cite{nakamura2022feminist}. Languages like French and Spanish have 
\emph{grammatical} gender unrelated to the speaker's gender~\cite{James1991spanish, Izabella2017french}.
In Thai, first-person pronouns differ by speaker gender~\cite{Saisuwan2015Thai}.
In Japanese, first-person pronouns like ``I'' and ``me'' 
are identity-laden: gender but also social identities like age, regional background, and social status~\cite{nakamura2007language,nakamura2022feminist}. Moreover, social and state structures enforce specific pronouns by gender~\cite{miyazaki2016japanese}. Thus, Japanese self-referents are a matter of \emph{intersectionality}, leading \citet{fujii2024silver} to coin the term 交差代名詞 (kousadaimeishi) for intersectional Japanese pronouns. 
One can \emph{avoid} self-referents to maintain identity privacy~\cite{kai2000,miyajima2018, Satoh2021Self, nakamura2022feminist}
or use the \emph{gender-neutral} pronoun 私 (watashi). Still, these approaches can be perceived as \emph{evasive}. Self-referents, with their grounding in identity exposure or evasion, are then linked to power.
Thus, it is crucial to analyze voice agent use of self-referents in an intersectional way, with sensitivity to social power and stereotypes~\cite{fujii2024silver, nakamura2020formation}. 
We must also confirm whether agents reinforce or challenge existing stereotypes. 
This topic may be specific to the Japanese context, but aligns with HCI initiatives on social identity~\cite{vieweg2015between} and intersectionality~\cite{kumar2019intersectional, schlesinger2017intersectional}.

Here, we investigated perceptions of ChatGPT voices when using various Japanese self-referents. We chose ChatGPT as a representative chatbot and the best platform to explore perceptions of AI due to its fluency~\cite{marchandot2023chatgpt}. We created videos that simulated conversations between ChatGPT and a user, featuring seven different self-referents and three different ChatGPT voices. Our research questions (RQs) were: \textbf{\emph{Can ChatGPT evade identity perceptions associated with voice through the use of neutral (RQ1) or elusive (RQ2) self-referents?}} We additionally considered the intersectional aspect in \textbf{\emph{RQ3: Can ChatGPT voices elicit other markers of identity through intersectional self-referents?}}
As the usage and meaning of terms can be confusing, we begin with definitions. RQ1 aims to \emph{evade}, i.e., avoid participant gendering to achieve gender \emph{neutrality} using neutral cues. The word \emph{evasive} will be used if no gendering happens by participants. RQ2 aims to \emph{elude}, i.e., allow for plurality in participant gendering to achieve gender ambiguity and/or queering expectations. The word \emph{elusive} will refer to situations where multiple forms of gendering happens within/across participants.
We found a powerful influence of voice on gender perceptions, the potential of certain self-referents to evade this effect, and intersectional personas that subtly emerge from specific combinations of self-referents and voices. We contribute the following: 

\begin{itemize}
    \item The first user perception results on ChatGPT voice, notably the Japanese versions of Juniper, Breeze, and Ember
    \item Empirical evidence that the combination of self-referent and voice creates a dynamic interaction among social identity markers, resulting in the emergence of new personas
    \item The first study to explore 
    gender-neutral self-referents, such as subject omission (No self-referent) and non-pronominal self-referents (NPSR), which are everyday and exploratory forms of self-reference in Japan
\end{itemize}

The paper title reflects the plurality that we explored here: ``inter alia'' means ``among other things,'' with the dual implication of exploring  intersectional and alternatives, or ``aliases,'' ``among'' the various modes of self-referent.
Our work sets the stage for future development and deployment of LLMs and text-to-speech (TTS) systems founded in empirical findings on the sociocultural phenomenon of social identity elicitation---and evasion---through Japanese self-referents and computer voice. Since LLMs and TTS systems can be deployed in various ``bodies''---from chatbots to social robots to voice assistants---this work traverses fields of study and may inspire similar work outside of the Japanese context.


\section{Theoretical Background}
\label{sec:background}

We now dive into how theories of sociolingiustics and social identity in the Japanese context can map onto computer agents. We focus on developing our hypotheses about self-referents and cite work on HCI and AI where possible.

\subsection{Gender Identity and Neutrality in Japanese Self-Referents (RQ1)}

The Japanese language offers a wide variety of first-person pronouns and self-referents. However, in using first-person pronouns, people are forced to share their identities, intents, and understandings of their relationship(s) with others~\cite{Satoh2021Self, nakamura2022feminist}. 
Previous work explored perceptions of ChatGPT~\cite{fujii2024silver}, but only considered \emph{pronouns} and not other forms of \emph{self-reference}. No known work has systematically investigated the range of Japanese self-referents. We add on by considering two key modes of self-reference in Japan: \emph{No self-referent} and \emph{NPSR} (described below). If most Japanese conversations occur without explicit subjects, this aspect should also be examined for chatbots~\cite{Lee2008overt}. 
This led us to ask:

\begin{quote}
    \textbf{RQ1:} Can ChatGPT voices elicit gender-neutral perceptions using ``neutral'' 
    modes of self-reference, specifically No self-referent, わたし (watashi), and non-pronominal self-reference (NPSR)?
\end{quote}


\subsubsection{No Self-Referent: Omission of the Subject}
In Japanese, the subject is often omitted when it is obvious or when objective situations are being described, such as in one-on-one conversations~\cite{kai2000, miyajima2018} 
In 330 minutes of face-to-face conversations between native Japanese speakers, about
84.5\% of possible first-person subjects were omitted~\cite{Lee2008overt}. This identity-evasive method can be used by any speaker in every situation. In the absence of a self-referent, voice may be influential~\cite{McGinn2019voice,Seaborn2021Voice}. 
Thus, we hypothesized:

\begin{quote}
    \textbf{H1:}
    In the case of no linguistic gender cue, the gender perception of ChatGPT will be based on the perceived voice gender.
\end{quote}

\subsubsection{わたし (Watashi): The Neutral Pronoun}
This pronoun is the most neutral and common method of self-reference~\cite{nakamura2014gender, nakamura2022feminist}, including for intelligent agents~\cite{fujii2024silver}. わたし (watashi) may be gender-neutral, as it is commonly used in a variety of situations by all people. Pepper, a robot designed to have no perceivable gender~\cite{pandey2018pepper}, also uses わたし (watashi).
However, 
わたし (watashi) is not completely gender neutral. All adults use わたし (watashi), but in private or in elementary school, わたし (watashi) is considered a pronoun for girls~\cite{nakamura2007language,nakamura2014gender,Satoh2021Self}. Engineers often attribute わたし (watashi) to their work to avoid gendering~\cite{pandey2018pepper}. However, \citet{fujii2024silver} showed that while わたし (watashi) can be perceived as neutral, it also elicits perceptions of femininity about 20\% of the time and masculinity 10\% of the time. 
When combined with a feminine voice (Juniper), the slight gender cues of わたし (watashi) may reinforce a feminine perception. 
We hypothesized:

\begin{quote}
    \textbf{H2-1:}
    In the case of わたし (watashi), the gender perception of ChatGPT will be based on the perceived voice gender.
\end{quote}
\begin{quote}
    \textbf{H2-2:} If Juniper (feminine) uses わたし (watashi), Juniper will be ascribed a feminine gender more frequently than the case of No self-referent (refer to H1).
\end{quote}
\begin{quote}
    \textbf{H2-3:} If Ember (masculine) uses わたし (watashi), perceptions of Ember’s gendering will be no different than the case of No self-referent (refer to H1).
\end{quote}

\subsubsection{Non-Pronominal Self-Reference (NPSR): Name as a Self-Referent}
NPSR or \emph{illeism} is when a person uses their own name in place of a self-referent. NPSR is linked to various stereotypes and social identities~\cite{Kojima2017npsr,takahashi2009okinawa, kajino2011,Satoh2021Self}. Names not socially or phonetically perceived as feminine or masculine may be deemed gender neutral or ambiguous. 
However, NPSR has various gender and age cues. Children and female students are known to use NPSR~\cite{Kawasaki2006name,Nishikawa2011name}. 
\citet{Nishikawa2011name} found that boys used their name but then switched to おれ (ore) or ぼく (boku) based on the context, while girls used their name or nickname or later transitioned to わたし (watashi).
\citet{Kojima2017npsr} reported that $\sim$11\% of female university students often used their first name or nickname
~\cite{miyazaki2016japanese, kajino2011,Kojima2017npsr,takahashi2009okinawa}. NPSR is also prominent in the media, used by female pop star Ayumi Hamasaki and writers of women's magazines~\cite{kajino2011}. NPSR is also used to build intimate relationships, as well as being an indicator of childishness, an aspect of desirable femininity in Japan~\cite{kajino2011, Kojima2017npsr}.

Voice 
can modify the effects of social cues~\cite{Seaborn2021Voice}. Voice can, for instance, work for or against appearance to shift gender perceptions~\cite{McGinn2019voice} and even override appearance, as with a gender-ambiguous voice given to binary-stereotyped robotic forms~\cite{torre2023ambig}. As such, despite the association with children and young women and the gender ambiguity of some names, we expect gendered perceptions of the ``neutral'' NSPR to align with the voice. We hypothesized:


\begin{quote}
    \textbf{H3-1:}
    In the case of NPSR, the gender perception of ChatGPT will be based on the perceived voice gender.
\end{quote}

Still, given the noted associations, we may expect more attributions of femininity when NPSR is used compared to a lack of self-referent, i.e., no cue whatsoever. We thus hypothesized: 

\begin{quote}
    \textbf{H3-2:} If Juniper (feminine) uses えーあい (AI) as the first-person pronoun, i.e., NPSR, Juniper will be ascribed a feminine gender more frequently than the case of No self-referent (refer to H1).
\end{quote}

No research has linked NPSR and masculinity (beyond childhood) in the Japanese context. In effect, NPSR could be neutralized when employed by a masculine voice, rendering it equal to a lack of self-referent.
We thus hypothesized:

\begin{quote}
    \textbf{H3-3:} If Ember (masculine) uses えーあい (AI) as the first-person pronoun, i.e., NPSR, perceptions of Ember’s gendering will be no different than the case of No self-referent (refer to H1).
\end{quote}

In the NSPR condition, we had ChatGPT use the name ``えーあい'' (e-ai, the Japanese pronunciation of ``AI'') to avoid any associations (and potential confounds) related to the ChatGPT brand. There are no Japanese names that sound like ``e-ai,'' so we are confident that there would be no impact on impressions, like participants having the same name.

\begin{figure*}[!ht]
\centering
\includegraphics[width=.85\textwidth]{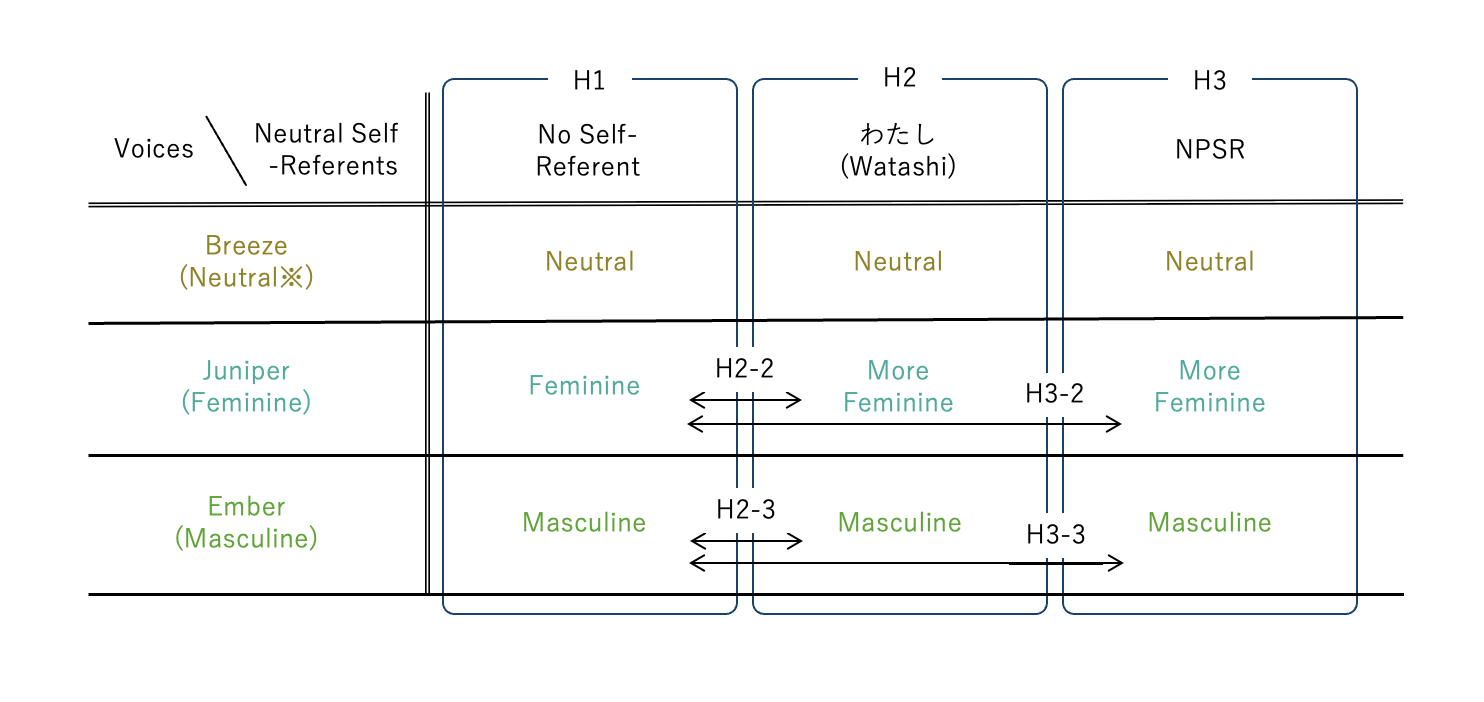}
\caption{Chart illustrating the hypotheses (H1-H3) for RQ1 that compare voices and self-referents.}
\Description{This table arranges three types of voices vertically and three neutral self-references horizontally. Three hypotheses are mapped to each self-referent, with four hypotheses comparing different self-referents within the same voice.}
\label{fig:RQ1chart}
\end{figure*}

\subsection{Gender Evasion by Queering Expectations (RQ2)}

\citet{fujii2024silver} showed that ChatGPT can easily evoke a persona in text by using pronouns. However, voice may be stronger than text, and other cues may disrupt typical perceptions~\cite{cameron2003language}. 
\emph{Queering} expectations 
may disrupt stereotypes 
and engender new identities~\cite{Light_2011queer}. ``Queer'' or ``gay'' language~\cite{cameron2003language}, referred to as おネエことば (onee-kotoba) in Japanese, is often labeled as deviant~\cite{maree2013onee}. Therefore, queering voices and self-referents might evoke the image of a persona with a specific sexuality or diversify identity perceptions, potentially reducing the power of hegemonic identities.
This may be an alternative to mixing and matching attributes to elicit novel perceptions~\cite{torre2023ambig}.
Also, the variety of pronouns in Japanese warrants a careful examination of effects and risks for ethical HCI. 
We therefore asked:

\begin{quote}
    \textbf{RQ2:} Can ChatGPT voices elicit gender-ambiguous perceptions by queering the use of intersectional first-person pronouns?
\end{quote}

\citet{fujii2024silver} showed that every pronoun elicited strong and diverse identity perceptions from ChatGPT. We select ぼく (boku) and じぶん (jibun) as masculine options and あたし (atashi) and わたくし (watakushi) as feminine options. 
While voice may not be gender-neutral~\cite{Seaborn_2022neutral,nass1997machines}, when we use a gender-neutral voice with gendered first-person pronouns, the pronoun gender cues may trigger gender perceptions of the AI. As such:

\begin{quote}
    \textbf{H4:} If ChatGPT uses Japanese feminine/masculine first-person pronouns with the gender-neutral Breeze voice, the pronouns will elicit feminine/masculine impressions.
\end{quote}

Celebrities, singers, and anime characters~\cite{Nishida2011bokukko} can influence girls to use ぼく (boku), a masculine pronoun~\cite{ide1997women, nakamura2014gender}. This indicates a deviation from gender roles~\cite{miyazaki2016japanese} and the expression of new values, which may be read as gender-neutral or ``cancelling'' gender through the mismatch~\cite{nakamura2014gender, miyazaki2016japanese}. The term ``ボクっ子／娘'' (ぼくっこ) (boku-kko), referring to a girl character usually played by a female voice actress who calls herself ぼく (boku), is an influential concept in Japanese subculture~\cite{Nishida2011bokukko}. 
ぼくっこ (boku-kko) creates a sense of competition among the various gender perceptions invoked and could be used to evade binary gendering. \citet{fujii2024silver} found that ぼく (boku) could elicit strongly masculine perception over 90\% from agents.

\begin{quote}
    \textbf{H5-1:} Using ぼく (boku) with feminine voices will elicit gender-ambiguous impressions of ChatGPT.
\end{quote}

じぶん (jibun), considered masculine, comes from military culture ~\cite{nakamura2014gender,kinuhata2007guntai}. \citet{fujii2024silver} found that じぶん (jibun) elicited a masculine impression $\sim$70\% of the time, but also elicited perceptions of femininity $\sim$5\% of the time, genderlessness $\sim$10\%, and ambiguity $\sim$15\% of the time. Nowadays, women students use じぶん (jibun) in formal situations~\cite{Ogino2007jibun}. In Okinawa, the usage rate is higher among women~\cite{takahashi2009okinawa}. Thus, we hypothesized:

\begin{quote}
    \textbf{H5-2:} Using じぶん (jibun) with feminine voices will elicit feminine impressions of ChatGPT.
\end{quote}

Sexual identities are often linked to specific speech styles, such as gay men using feminine language, especially pronouns and 文末詞 (bunmatsu-shi, or sentence-ending words)~\cite{kawano2016onee, maree2013onee, nakamura2022feminist}. 
In Japan, this type of speech is known as おネエ言葉 (onee-kotoba) and is used theatrically by queer celebrities on Japanese TV shows~\cite{maree2007language, maree2013onee, kawano2016onee}.  
This reflects heteronormativity by framing homosexuality as gender deviance. Use of あたし (atashi) as おネエ言葉 (onee-kotoba) risks eliciting stereotypes about queer people that are already perpetuated through the media ~\cite{kawano2016onee, nakamura2022feminist}. 
While taking such concerns into consideration, we will also examine how, similar to Juniper using ぼく (boku), Ember using あたし (atashi) creates a sense of competition among the various gender perceptions invoked and could be used to evade binary gendering.
We hypothesized:

\begin{quote}
    \textbf{H6-1:} Using あたし (atashi) with masculine voices will elicit gender-ambiguous impressions of ChatGPT.
\end{quote}

All adults use わたくし (watakushi)~\cite{nakamura2014gender, nakamura2022feminist}, including men in speeches~\cite{nakamura2014gender, nakamura2022feminist}. However, it was perceived as feminine and elusive for text-based ChatGPT~\cite{fujii2024silver}, possibly because of its ``feminine''' politeness. We hypothesized:

\begin{quote}
    \textbf{H6-2:} Using わたくし (watakushi) with masculine voices will elicit masculine impressions of ChatGPT.
\end{quote}

\begin{figure*}[tbh]
\centering
\includegraphics[width=.95\linewidth]{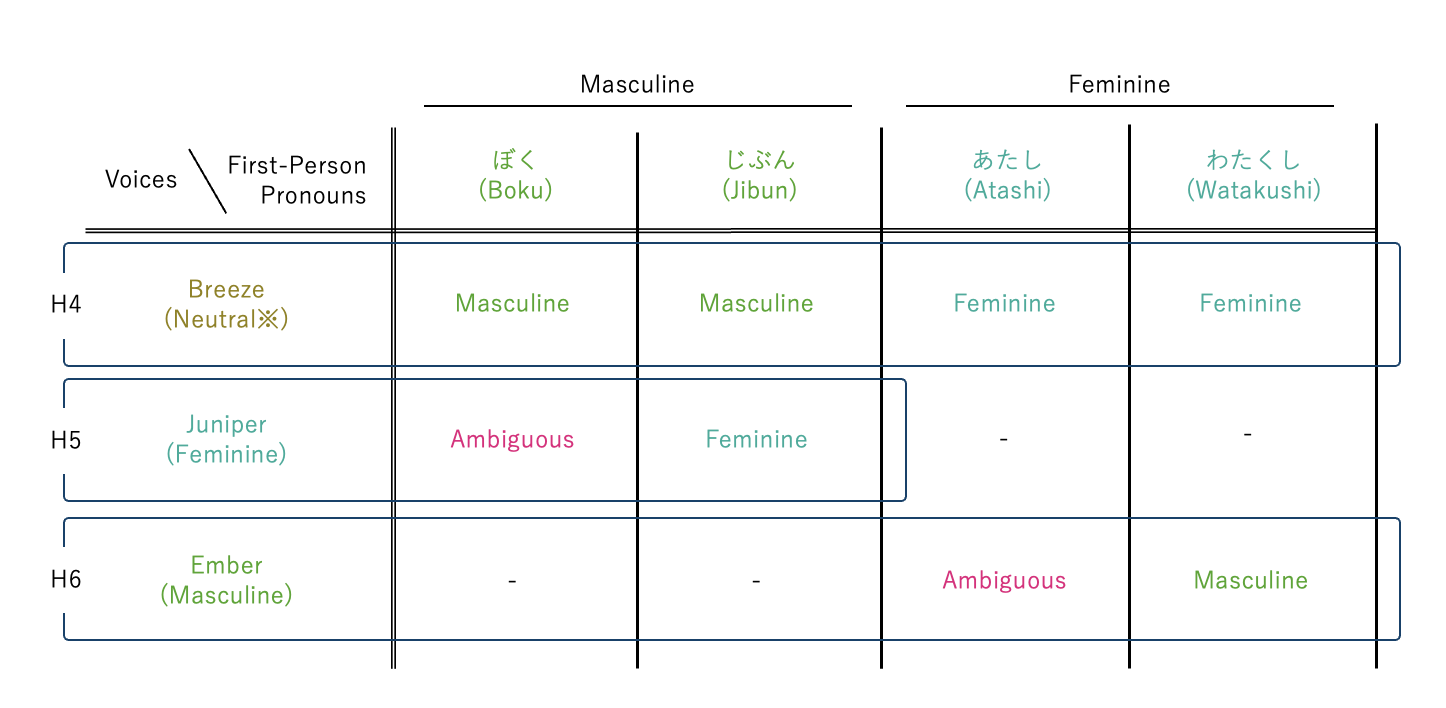}
\caption{Chart illustrating the hypotheses (H4-H6) for RQ2 that compare voices and pronouns.}
\Description{This chart maps out three hypotheses by arranging the types of voices vertically and the Japanese first-person pronouns horizontally. The four missing cells represent combinations where the gender cues of the voices and pronouns overlap, and thus were not tested in this study.}
\label{fig:RQ2chart}
\end{figure*}

\subsection{Intersectionality and Identity in Japanese Self-Referents (RQ3)}
\label{sec:bgintersectional}

Japanese self-referents evoke complex identities: not only gender but also age, region, and social status. When an artificial agent uses Japanese first-person pronouns, these usually need to be interpreted as intersectional. For example, \citet{fujii2024silver} discovered that when ChatGPT used the pronoun うち (uchi), this evoked not only a feminine perception but also conveyed the image of a young and regional persona. Thus, we consider the intersectional aspects of the self-referents introduced in RQ2. 
Our guiding RQ was:

\begin{quote}
    \textbf{RQ3: Can ChatGPT voices elicit other markers of identity through intersectional self-referents?} 
\end{quote}

\subsubsection{Agedness}
In Japan, children who have not yet mastered the correct use of first-person pronouns 
often use NPSR when referring to themselves~\cite{Kawasaki2006name}. Additionally, it is common for teenagers to use NPSR when speaking to close friends or family members~\cite{miyazaki2016japanese, kajino2011,Kojima2017npsr,takahashi2009okinawa}. While there are no studies in the field of computing that address NPSR, based on prior research in Japanese sociolinguistics, we hypothesized: 

\begin{quote}
    \textbf{H7:} In the case of NPSR, the age perception of all voices will be younger than when each voice is using わたし (watashi) or No self-referent. 
\end{quote}

\citet{fujii2024silver} found that ぼく (boku), じぶん (jibun), and あたし (atashi) elicited a younger impression than 私 (watashi-kanji; same pronunciation as わたし or watashi) for the text-based version of ChatGPT. As such, we hypothesized:

\begin{quote}
    \textbf{H8-1,2,3:} The age perception of Breeze and Juniper using ぼく (boku) (H8-1) and じぶん (jibun) (H8-2) and Breeze and Ember using あたし (atashi) (H8-3) will be younger than わたし (watashi), which is the most common pronoun. 
\end{quote}

\citet{fujii2024silver} found that わたくし (watakushi) elicited an adult or middle-aged impression for the text-based version of ChatGPT. As such, we hypothesized for voice: 

\begin{quote}
    \textbf{H8-4:} The age perception of Breeze and Ember using わたくし (watakushi) will be older than わたし (watashi), which is the most common pronoun. 
\end{quote}

\subsubsection{Kawaiiness}

``Kawaii'' is a uniquely Japanese concept similar to ``cute,'' which has spread globally through various media and anime culture~\cite{Nittono2010behavioral}. Kawaii has complex meanings: not only cute, but also lovable, pitiable, small, weak, and giving off the feeling of ``needing protection.'' Research on kawaii in HCI and HAI has begun~\cite{wang2024kawaii}. \citet{seaborn_can_2023} found a negative correlation between the perceived cuteness of different computer voices and age perceptions, indicating that kawaii is associated with youthfulness. 
The girlishness of NPSR, which implies femininity (refer to H3-2) and youthfulness (H7), may lead to perceptions of kawaii. The concept of a ``ぼくっこ'' (boku-kko), or a girl who uses the masculine pronoun ぼく (boku), was originally nurtured within kawaii Japanese animation culture~\cite{Nishida2011bokukko}. So, the use of ぼく (boku) by Juniper should evoke perceptions of kawaii. We hypothesized: 

\begin{quote}
    \textbf{H9-1:} Juniper using NPSR will be perceived as more kawaii than Juniper using わたし (watashi), No self-referent, and じぶん (jibun).
\end{quote}
\begin{quote}
    \textbf{H9-2:} Juniper using ぼく (boku) will be perceived as more kawaii than Juniper using わたし (watashi), No self-referent, and じぶん (jibun).
\end{quote}

\subsubsection{Ruralness and Formality}
Regional dialects, often used in private settings, are frequently viewed as inferior to the national ``standard language,'' which is associated with public settings~\cite{takagi2005kansai}. People may feel ashamed of their dialect or accents and conceal them, especially women~\cite{kumagai2010dialect, nakamura2014gender}. However, dialects shape regional identities and intersect with other social identities~\cite{takagi2005kansai, kumagai2010dialect, nakamura2014gender}. 
\citet{fujii2024silver} found that じぶん (jibun) elicited a rural region impression (known for its unique dialect) in text-based ChatGPT.  じぶん (jibun) is used throughout Japan, but in Kansai, it can also be used as a second-person pronoun, which will elicit a regional or dialectal image~\cite{Kigawa2011jibun, muranaka2015dialect}. As such, we hypothesized: 

\begin{quote}
    \textbf{H10-1:} Juniper and Breeze using じぶん (jibun) will elicit a rural image compared to other pronouns.
\end{quote}

わたくし (watakushi) is the most formal first-person pronoun used in everyday situations, regardless of gender. However, it is usually used in public speeches and is often associated with the upper class~\cite{kumadaki2006language, ide1990and}. \citet{fujii2024silver} also found that わたくし (watakushi) elicited a formal impression, where participants used words like ``elegant'' and ``polite'' to describe the text-based version of ChatGPT. We therefore hypothesized:

\begin{quote}
    \textbf{H10-2:} Ember and Breeze using わたくし (watakushi) will elicit a formal impression compared to each voice using other pronouns.
\end{quote}

\begin{figure*}[!ht]
\centering
\includegraphics[width=\textwidth]{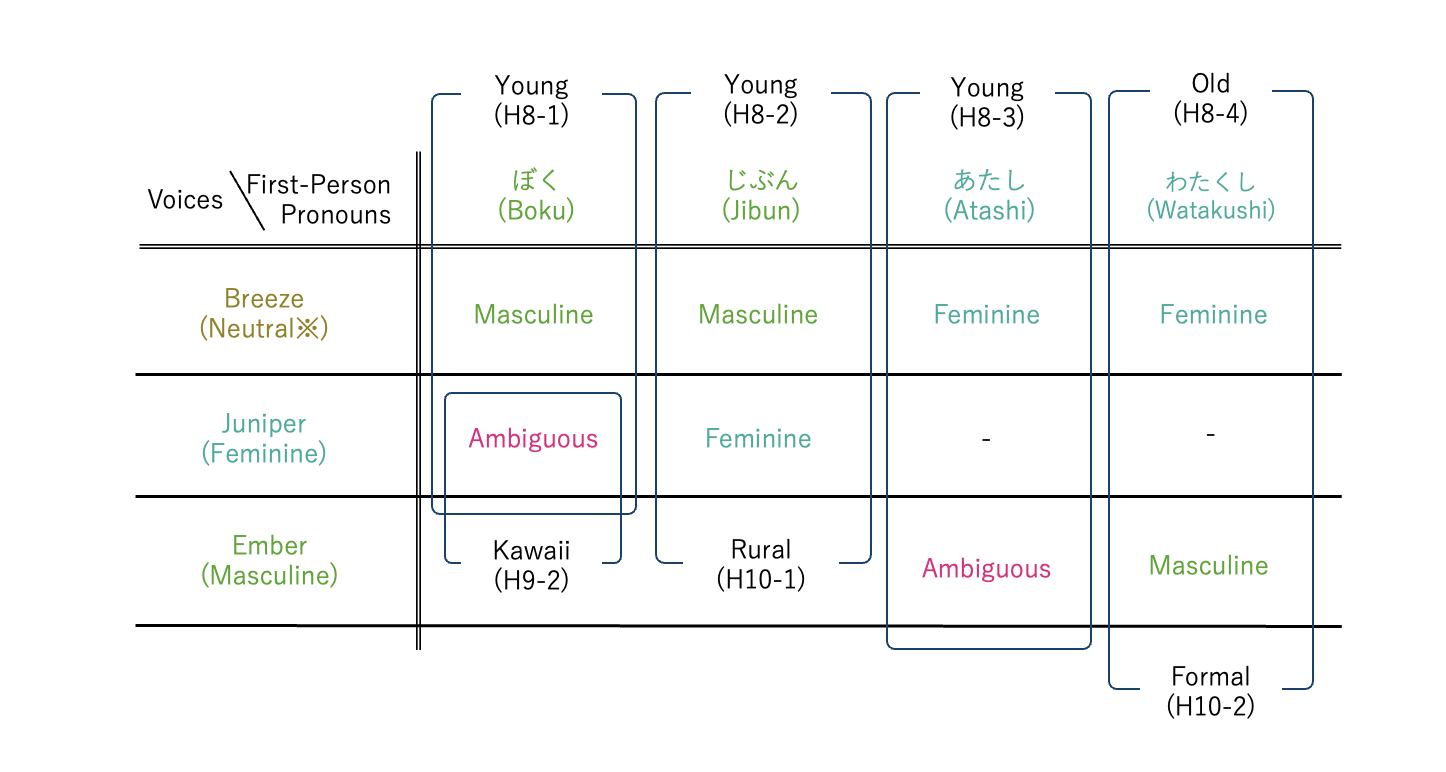}
\caption{Chart illustrating the hypotheses (H7-10) for RQ3 that compare voices and pronouns.}
\Description{This chart maps out three hypotheses by arranging the types of voices vertically and the Japanese first-person pronouns horizontally. The four missing cells represent combinations where the gender cues of the voices and pronouns overlap, and thus were not tested in this study.}
\label{fig:RQ3chart}
\end{figure*}


\section{Methods}
\label{sec:methods}
We carried out a within-subjects online user perceptions study. All participants experienced all combinations of voice and self-referent in a random order. We replicated the protocol of \citet{fujii2024silver}, with three exceptions: (i) we added ChatGPT voice to the video stimuli; (ii) we only used one of the confirmed neutral conversation scripts; and (iii) we did not target specific regions of Japan when recruiting. Our protocol was registered before data collection on July 31\textsuperscript{st}, 2024\footnote{\url{https://osf.io/8pngr}}; however, we updated it on August 28\textsuperscript{th} to add RQ3 and related hypotheses (before data analysis but after data collection and basic processing)\footnote{We updated it again on August 31\textsuperscript{st} after noticing a typo where we used the wrong voice names in the hypotheses.}. This research was approved by the institutional ethics board (\#2023081).

\subsection{Participants}
Recruitment was via Yahoo! Crowdsourcing Japan, an online recruitment platform similar to Prolific and Amazon Mechanical Turk that relies on identity verification for basic demographics\footnote{\url{http://crowdsourcing.yahoo.co.jp/}}. 166 people were recruited from the general pool and a further 52 women were recruited to correct for oversampling of men. 
The data of thirteen inattentive participants was removed for 204 participants total, which were 89 (43.6\%) women, 110 (53.9\%) men, 4 (2.0\%) no answer, and 1 (0.5\%) other. 
Compensation was 1200 yen/hour. 

\label{section:procedure}
\subsection{Procedure}
Participants were given a link to the questionnaire on SurveyMonkey\footnote{\url{https://www.surveymonkey.com/}}. They provided consent on the first page. The next page was an attention check, where they watched a video of a sample conversation without voice or self-referents, and then input the number that appeared in the video. Then, 17 combinations self-referents and voices were presented in a random order. On each page, they watched a short video depicting an interaction with ChatGPT using a specific self-referent. They then provided their impressions of the ``AI'' in rating scales and open-ended items. Afterwards, they provided demographics and received a Yahoo! payment code. The study took $\sim$20 minutes.

\subsection{Materials}
\begin{figure*}[!t]
\centering
\begin{subfigure}[t]{0.49\linewidth}
    \centering
    \includegraphics[width=.95\linewidth]{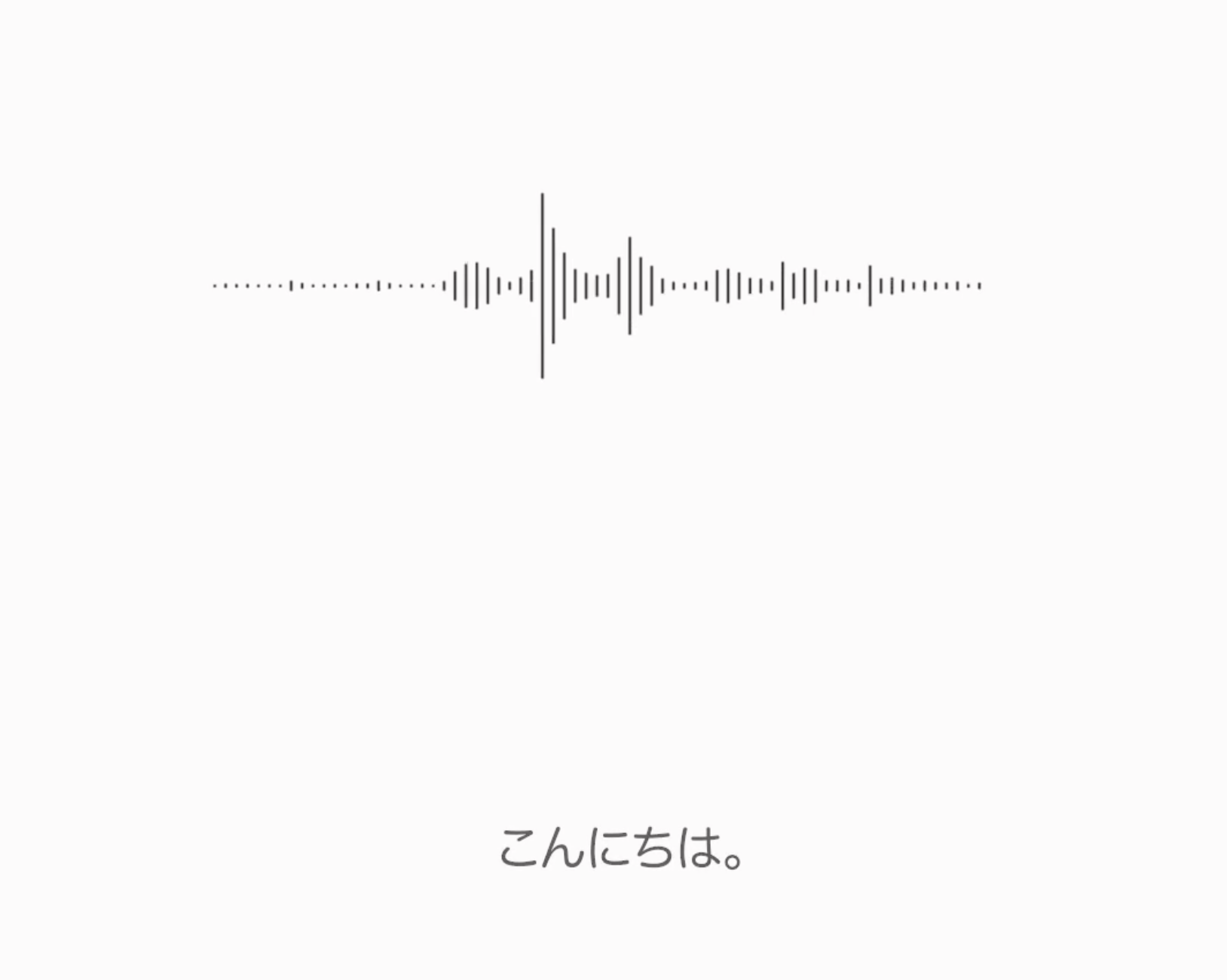}
    \caption{Screen shot of the video clip with ChatGPT voice.}
    \Description{This scene depicts a user greeting ChatGPT with ``こんにちは (Hello).'' The AI responds with voice. The text fades in from the bottom of the screen, and as the AI's voice plays, the waveform undulates in sync with the audio.}
    \label{fig:videoclip}
\end{subfigure}%
    ~ 
\begin{subfigure}[t]{0.49\linewidth}
    \centering
    \includegraphics[width=.95\linewidth]{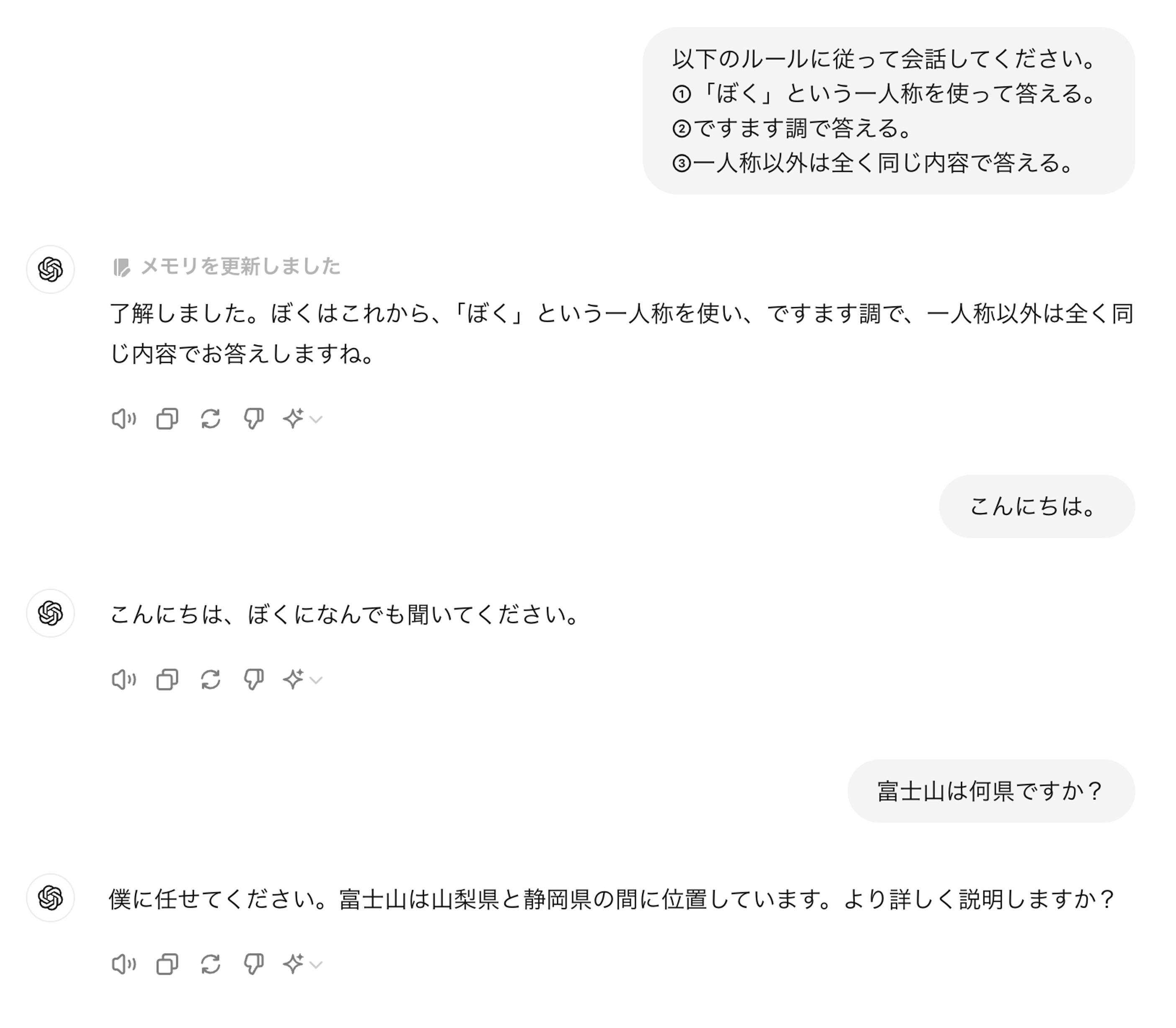}
    \caption{Prompt screen in the ChatGPT interface.}
    \Description{This depicts the instructions given to ChatGPT and the conversation for recording the voice.}
    \label{fig:prompt}
\end{subfigure}

\caption{Illustration of the stimuli generation process with ChatGPT.}
\label{fig:generationprocess}
\end{figure*}

We created 17 videos, each $\sim$15 seconds in length, designed for each hypothesis. 
First, we initiated and video-recorded a text-based conversation with ChatGPT about the confirmed neutral topic of Mount Fuji~\cite{fujii2024silver}. 
We entered the following prompts into the ChatGPT front-end interface (\autoref{fig:generationprocess}) to: set and enforce use of (i) ChatGPT’s self-reference mode; (ii) the です (desu) and ます (masu) sentence ending structure; and (iii) the same content except the self-reference. For Prompt (ii), です (desu) and ます (masu) are neutral and do not evoke social identities~\cite{kobayashi2013language}. 

\begin{itemize}
    \item {Please follow these rules when we are conversing. \\
    JP: 以下のルールに従って会話してください。}
    \item {Prompt (i): Use the pronouns [watashi/AI/boku/jibun/atashi/ watakushi] when you respond.\\
    JP: 一人称を「わたし/えーあい/ぼく/じぶん/あたし/わたくし」にして答える。}
    \begin{itemize}
    \item {Prompt (i, for No self-referent): Do not use any first-person pronouns when you respond.\\
    JP: 一人称を用いずに答える。}
    \end{itemize}
    \item {Prompt (ii): End your sentences with ``desu/masu.''\\
    JP: ですます調で答える。}
    \item {Prompt(iii): Answer with exactly the same content, excepting the first-person pronouns.\\
    JP: 一人称以外は全く同じ内容で答える。}
\end{itemize}

Videos were modified in Adobe Premiere Pro and After Effects to remove visual markers of ChatGPT, to avoid the ChatGPT brand as a confound or prime~\cite{head1988priming}. A simple monochrome visualization comprised of a waveform and text was created (\autoref{fig:generationprocess}).
We then generated and added ChatGPT audio to the videos in post-production. 
Videos were uploaded to YouTube and embedded into the questionnaire. These  can be accessed on OSF\footnote{\url{https://osf.io/8pngr}}.
ChatGPT voice depends on the version and language setting. In July 2024, we had five voices: Sky (suspended), Juniper, Breeze, Ember, Cove (in order of voice pitch). We chose Juniper (as feminine), Breeze (as gender-neutral), and Ember (as masculine).

\subsection{Instruments and Measures}

\subsubsection{Genderedness}
We used the nominal ``gender-expansive'' options from the text-only study~\cite{fujii2024silver}, originally from \citet{Seaborn2022Expansive}. 
The item was: ``What gender is this AI, do you feel? Please select the closest option.'' (このAIの性別・ジェンダーはどのように感じられましたか? 最も近いものを一つ選択してください) 
Response options were:
feminine (女性的),
masculine (男性的),
gender-ambiguous (a mixture of feminine and masculine characteristics, 女性的でも男性的でもある), and
genderless (女性的でも男性的でもない: 性別を感じない), plus an
optional free text field called ``another (gender)'' (その他) for alternative categorizations. 
As per \citet{fujii2024silver}, we use the term ``gender elusive'' to encompass both gender ambiguous (indeterminate) and gender neutral attributions (genderless). These represent indeterminate gender neutral perceptions: unbiased or undecided (ambiguous) and free of perceivable genders (genderless).

\subsubsection{Agedness}
Age was used to examine the relationship between gender, age, and kawaii perceptions. 
We used the nominal age options from voice-based CUI work~\cite{seaborn_can_2023}. 
The item was: ``How old do you feel this AI is? Please select the closest option.'' (このAIは何歳くらいだと感じましたか? 最も近い選択肢を一つ選んでください) Response options included: baby (赤ちゃん 0-2歳), child (子ども3-12歳), teenager (ティーンエイジャー13-19歳), adult (大人20-39歳), middle-aged (中年40-64歳), older adult (高齢者65歳以上), and ageless (年齢は特定できない). We also provided an ``another (age category)'' (その他) free text field. 

\subsubsection{Kawaiiness}
We used the 5-point Likert scale item from \citet{seaborn_can_2023}: 
``This AI feels kawaii.'' (このAIは可愛いと感じる) with the response options 1: Strongly Disagree, 2: Disagree, 3: Neither, 4: Agree, 5: Strongly Agree (全く同意しない〜あまり同意しない〜どちらとも言えない〜やや同意する〜強く同意する).

\subsubsection{Formality, Region, and ぼくっこ (boku-kko)}
We used an open field asking ``What kind of persona comes to mind when you think about this AI?'' (このAIについて、どんな人物像を思い浮かべますか?) to 
complement the quantitative measures.

\subsubsection{Complementary Measures and Distractors}
We added items on Anthropomorphism/Humanlikeness as distractors 
and to gather insights on the identity attributions. 
We used the 5-point Likert scale items from~\citet{baird_perception_2018} and divided the poles (humanlike and artificial) into two items to match the structure of the other items in the questionnaire~\cite{seaborn_can_2023}. 
The items were: ``This AI feels humanlike'' (このAIは人間的らしい), ``This AI feels machinelike'' (このAIは機械的だ), 
``This AI feels reliable'' (このAIは信頼できる), ``This AI feels competent'' (このAIは有能だ), ``This AI feels approachable'' (このAIは親しみやすい), and ``This AI feels like a Japanese native speaker'' (このAIは日本語を母語としている). Each item used the same scale as for kawaiiness.

\subsection{Data Analysis}
We carried out quantitative and qualitative analyses.

\subsubsection{Main Quantitative Analyses}
We used Google Sheets for descriptive statistics and IBM SPSS Statistics 29 for Friedman tests and to test statistical assumptions, e.g., sphericity. When test assumptions were met, we used the R-based online Statistics Kingdom\footnote{\url{https://www.statskingdom.com/}} calculator for the Chi-square tests; otherwise, tests were run using equation~\ref{eq:1} in Microsoft Excel, with $p$-values calculated using the Social Science Statistics online calculator~\cite{chiSquareCalc}. We generated counts for categorical data and means for Likert data (in the Supplementary Material).
Chi-squared tests were used to compare predicted distributions of nominal categories to, in most cases, the distribution of the sum of No self-referent, NPSR, and わたし (watashi), referred to as neutral. 
This required multiple comparisons for each hypothesis, so Bonferroni corrections were applied. Due to the $\chi{^2}$ formula, expected values that were zero were removed, notably ``other.''
Also, the baby and child categorizations were combined into one category due to frequent 0 counts. To allow for unequal sample sizes, the expected values were derived from percentages. As such, the formula $\chi{^2}$ was:

\begin{equation}~\label{eq:1}
    \chi{^2} = \sum\frac{(o_i-e_i)^2}{e_i}
\end{equation}
where $e_i = \frac{n(\%_i)}{100}$,

$o_i =$ the number of observations,

$n$ is the sample size of the comparator distribution.

Effect sizes ($w$) for each Chi-square test were calculated ($0.1$: small, $0.3$: medium, $0.5$: large~\cite{effectSize}\footnote{Some \textit{w} values may be greater than 1 due to the formula allowing for any score above zero.}). Some expected values were very small, making the test more sensitive to differences and possibly overestimating significance. As such, some results are indicative.

\subsubsection{Supplementary Quantitative Analyses}

People may consider a voice to be a singular entity when it is presented across multiple bodies~\cite{luria2019re}. As such, we analyzed observed variance in voice gendering within participants as a measure of whether individuals were affected by differences in self-referent for a single voice, regardless of presentation order. First, 
we relabelled the data such that participants who considered the voice masculine were scored 1, feminine was 2, ambiguous was 3, genderless was 4 and another gender was 5.
We then calculated the standard deviation (SD) 
for each participant by voice and self-referent. 
An SD of 0 
indicates consistent attribution of the same gender category to the voice, regardless of pronoun. An SD greater than 0 indicates attributions varied by gender category for a given voice \emph{at least once}. As the numbers assigned to each gender category were arbitrary, we did not consider SD scores beyond 0 or more than 0.

\subsubsection{Qualitative Analysis}
We used a hybrid thematic analysis approach~\cite{Proudfoot_2022}. First, a Japanese native researcher reviewed the dataset, 
categorizing essential verbs and adjectives
. The author then deductively applied the thematic framework from the text-based ChatGPT research~\cite{fujii2024silver}, notably formality and region. The data was repeatedly tagged, the codes refined until theoretical saturation was achieved. 
A second Japanese native researcher independently categorized 25\% of the data, inductively adding analytic codes for kawaii and ぼくっこ (boku-kko). Inter-rater reliability (IRR) was assessed with the Kappa statistic~\cite{julius2005kappa}. A minimum value of 0.89 was confirmed for each.


\section{Results}
The full descriptives for each voice can be found in the Supplementary Materials. 
Visualisations for genderedness (\autoref{fig:RQ1result} and \autoref{fig:RQ2result}) and agedness (\autoref{fig:RQ3result_age}) are presented. 
The results of the thematic analysis, and code counts for region, formality, ぼくっこ (boku-kko) can be found in \autoref{table:thematic}.
\subsection{Gender Identity and Neutrality in Japanese Self-Referents (RQ1)}

The expected values were the chance that each gender (masculine, feminine, ambiguous, or genderless) would be chosen at random (i.e., 25\%) compared to the observed categorisations when no-pronoun, わたし (watashi), and NPSR were used by each voice. 
A Bonferroni correction of 9 was applied.
A Chi-squared test indicated a statistically significant difference in the gendering of Juniper with all three pronouns compared to chance (all $\chi^2(3, 204) >300, p<.001, w>1$), where Juniper was categorised as feminine most frequently (Table~\ref{table:h1ChiSquare}). There were also statistically significant differences for Breeze ($\chi^2(3, 204) >400, p<.001, w>1$), and Ember ($\chi^2(3, 204$\footnote{$n$ was 203 for わたし (watashi).}) $>550, p<.001, w>1$). Both voices were categorised most frequently as masculine.
As such, we can answer the hypotheses as follows:

\begin{quote}
    \textbf{H1:} In the case of no linguistic gender cue, the gender perception of ChatGPT will be based on the perceived voice gender.\\
    $\rightarrow$ \textbf{Supported: Gender via voice}
\end{quote}
\begin{quote}
    \textbf{H2-1:}
    In the case of わたし (watashi), the gender perception of ChatGPT will be based on the perceived voice gender.\\
    $\rightarrow$ \textbf{Supported: Gender via voice}
\end{quote}
\begin{quote}
    \textbf{H3-1:}
    In the case of NPSR, the gender perception of ChatGPT will be based on the perceived voice gender.\\
    $\rightarrow$ \textbf{Supported: Gender via voice}
\end{quote}

\begin{table*}[!ht]
\caption{Chi-squared test results by voice, comparing No self-referent, NPSR, and わたし (watashi) to genderedness (masculine, feminine, ambiguous, genderless).
Juniper with No self-referent was then compared to NPSR or わたし (watashi). *$p <.05$, *** $p <.001$.}
\label{table:h1ChiSquare}
\begin{tabular}{lllrrrrr}
\toprule
Assumed distribution & Voice   & Comparator & $\chi^2$& Adj. $p$  & Sig.        & Comparator $n$ & Effect size ($w$) \\ \midrule
Chance               & Juniper & No Self-Referent        & 429.80                 & $<.001$ & *** & 204          & 1.45            \\
 &         & わたし (Watashi)    & 337.76                 & $<.001$ & *** & 204          & 1.29            \\
 &         & NPSR       & 396.35                 & $<.001$ & *** & 204          & 1.39            \\ \cline{2-8} 
 & Breeze  & No Self-Referent         & 520.51                 & $<.001$ & *** & 204          & 1.60            \\
 &         & わたし (Watashi)    & 491.65                 & $<.001$ & *** & 204          & 1.55            \\
 &         & NPSR       & 424.24                 & $<.001$ & *** & 204          & 1.44            \\ \cline{2-8} 
 & Ember   & No Self-Referent         & 604.04                 & $<.001$ & *** & 204          & 1.72            \\
 &         & わたし (Watashi)    & 593.16                 & $<.001$ & *** & 203          & 1.71            \\
 &         & NPSR       & 572.75                 & $<.001$ & *** & 204          & 1.68            \\ \hline
 No Self-Referent & Juniper & わたし (Watashi) & 12.52 & .012 & * & 204 & 0.25 \\
  &    & NPSR & 3.83 & .562 &  & 204 & 0.14 \\
 \bottomrule
\end{tabular}
\end{table*}

We expected that Breeze would be deemed gender-neutral or ambiguous, as described across media sources. However,  Breeze was perceived as masculine by most, so we did not use it as ``gender-neutral'' in our analyses.

We next compared the gender distribution for No self-referent and わたし (watashi) for Juniper. 
This was run alongside testing H3-2 and so a Bonferroni correction of 2 was applied. There was a statistically significant difference (\autoref{table:h1ChiSquare}) in the distributions, whereby No self-referent was significantly more feminine than わたし (watashi), which was more ambiguous and masculine.

\begin{quote}
    \textbf{H2-2:} If Juniper (feminine) uses わたし (watashi), Juniper will be ascribed a feminine gender more frequently than the case of No self-referent (refer to H1).\\
    $\rightarrow$ \textbf{Rejected: Opposite found}
\end{quote}

\begin{figure*}[!ht]
\centering
\includegraphics[width=\textwidth]{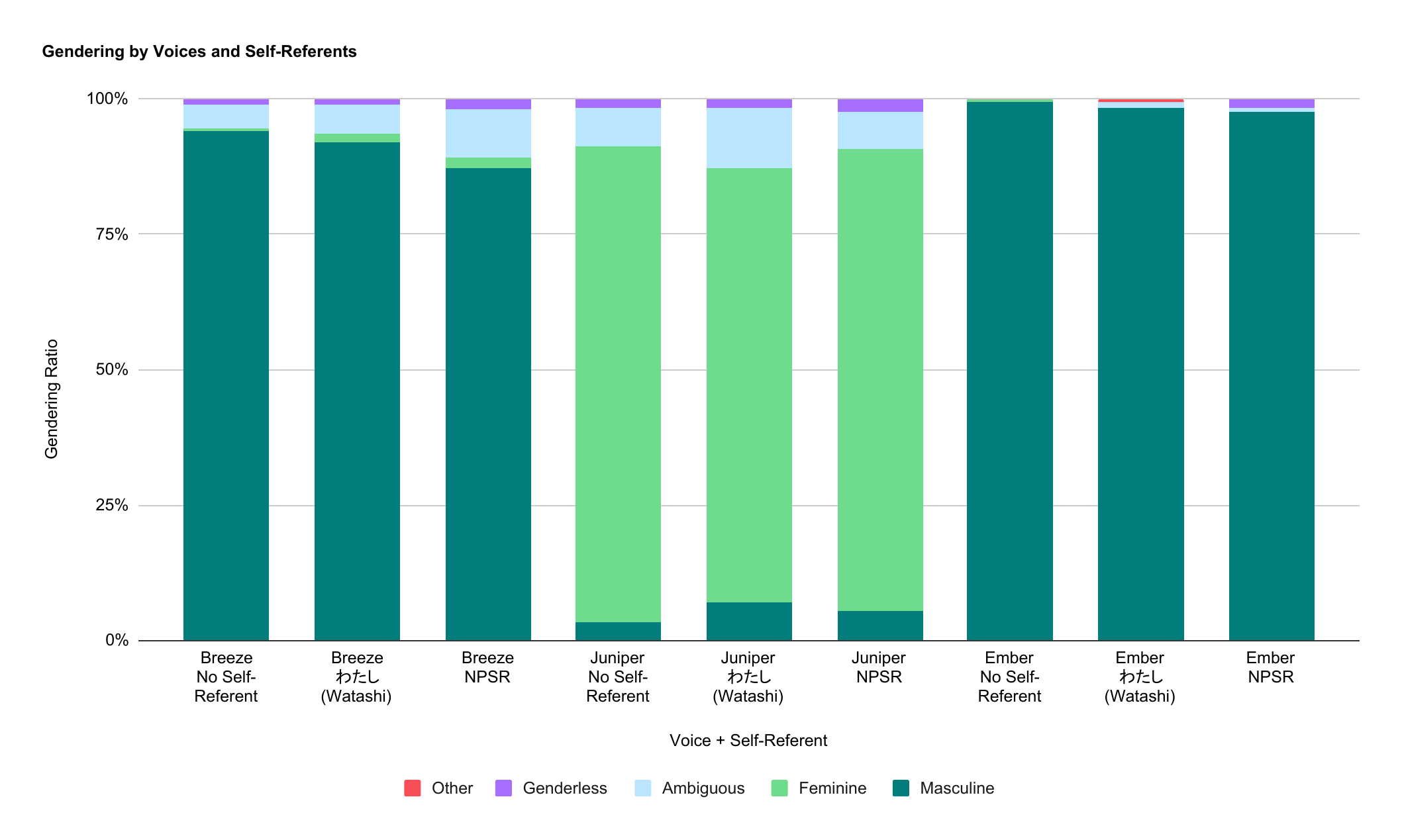}
\caption{Gender categorization for each voice using each self-referent as a percentage ($N=204$).}
\Description{All voices using each self-referent and their gender attributions, including masculine, feminine, ambiguous, genderless, and other. Breeze had about 90\% associations with masculine and about 7\% with ambiguous. Juniper had over 80\% associations with feminine and 10\% with ambiguous. Ember had about 99\% association with masculine. }
\label{fig:RQ1result}
\end{figure*}

Both Ember using わたし (watashi) and Ember not using a cue had virtually 100\% masculine attributions: 201 and 203 of $N=204$, respectively. The data thus did not meet the criteria to run a statistical test, such as the Chi-square. Nevertheless, the descriptive statistics speak for themselves (refer to \autoref{fig:RQ1result}). As such:

\begin{quote}
    \textbf{H2-3:} If Ember (masculine) uses わたし (watashi), perceptions of Ember’s gendering will be no different than the case of No self-referent (refer to H1).\\
    $\rightarrow$ \textbf{Supported: Masculine via voice}
\end{quote}

Alongside H2-2, we also investigated No self-referent and NPSR for Juniper. Again, a Bonferroni correction of 2 was applied. There was no statistically significant difference (\autoref{table:h1ChiSquare}).

\begin{quote}
    \textbf{H3-2:} If Juniper (feminine) uses えーあい (AI) as the first-person pronoun, i.e., NPSR, Juniper will be ascribed a feminine gender more frequently than the case of No self-referent (refer to H1).\\
    $\rightarrow$ \textbf{Rejected: No difference}
\end{quote}

The result found for Ember using わたし (watashi) was the same for NPSR: 199 of 204 counts for masculine. Thus:

\begin{quote}
    \textbf{H3-3:} If Ember (masculine) uses えーあい (AI) as the first-person pronoun, i.e., NPSR, perceptions of Ember’s gendering will be no different than the case of the case of No self-referent (refer to H1).\\
    $\rightarrow$ \textbf{Supported: Masculine via voice}
\end{quote}

\subsection{Gender Evasion by Queering Expectations (RQ2)}

\hl{Since Breeze was not perceived as gender-neutral, we could not answer the hypothesis:}

\begin{quote}
    \textbf{H4:} If ChatGPT uses Japanese feminine/masculine first-person pronouns with the gender-neutral Breeze, the pronouns will elicit feminine/masculine impressions.\\
    $\rightarrow$ \textbf{\hl{Cannot answer: Breeze not gender-neutral}}
\end{quote}

\begin{figure*}[!ht]
\centering
\includegraphics[width=\textwidth]{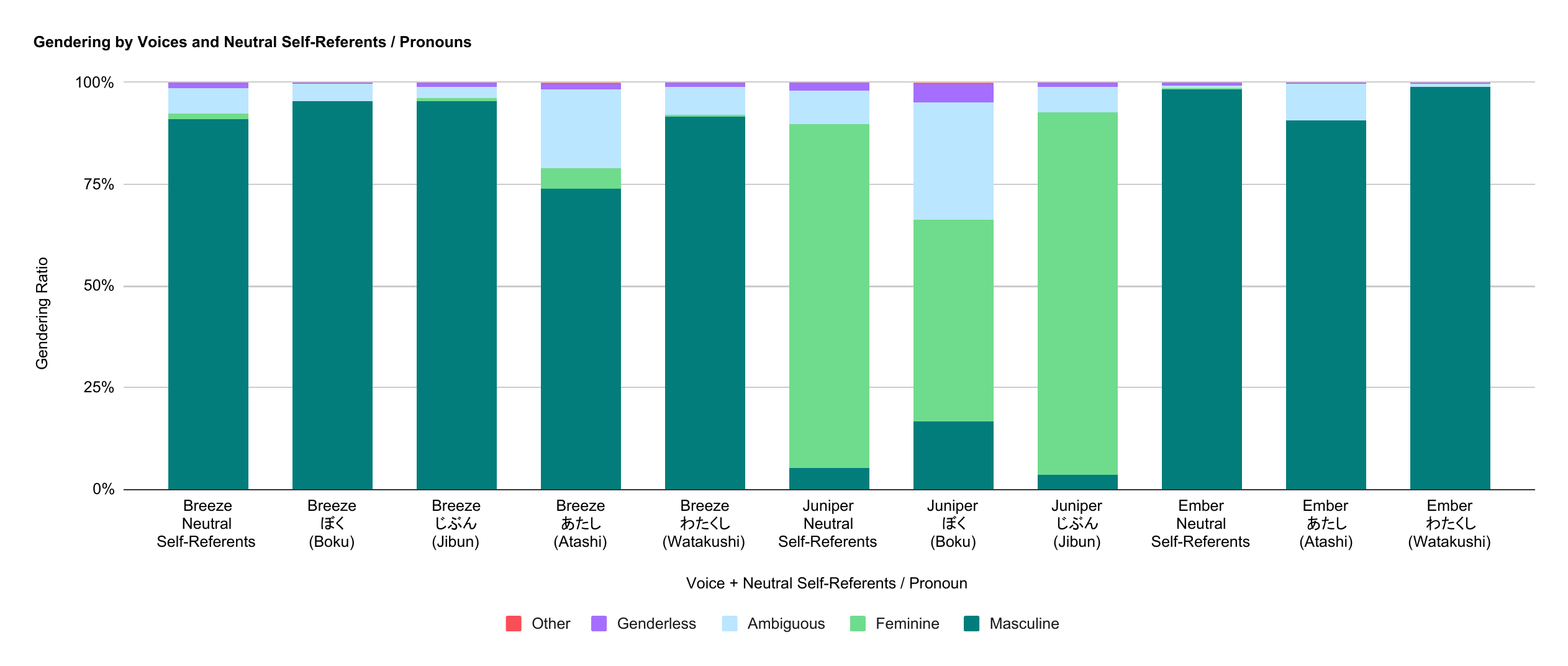}
\caption{The gender categorization of each voices using each self-referents or pronouns as a percentage ($N=204$).}
\Description{All voices using each self-referents or pronouns and their assignations of gender, including masculine, feminine, ambiguous, genderless, and other. Breeze using neutral self-referents, ぼく (boku), and じぶん (jibun)  had over 90\% associations with masculine, and using あたし (atashi) had about 70\% with masculine and 20\% with ambiguous. Juniper using neutral self-referents and じぶん (jibun) had about 90\% associations with feminine and, using ぼく (boku) had about 50\% with feminine and 30\% with ambiguous. Ember had over 90\% association with masculine. }
\label{fig:RQ2result}
\end{figure*}

Despite this, we did conduct an exploratory analysis to investigate the effect of the pronouns with this voice.
We compared the distributions of gender categorisations for the neutral pronouns compared to ぼく (boku), じぶん (jibun), あたし (atashi), and わたくし (watakushi). A Bonferroni correction of 4 was applied, and the expected values were calculated as per \autoref{eq:1}, to account for the different sample sizes.
The results (\autoref{tab:BreezeGenderComp}) indicated that gender perceptions of Breeze only differed when あたし (atashi) was used, whereby the voice was perceived as less masculine and more feminine or ambiguous (\autoref{fig:RQ2result}). Breeze using あたし (atashi) could then be gender evasive. 

\begin{table*}[!th]
\caption{Chi-squared test results comparing the gender perception of Breeze when using ぼく (boku), じぶん (jibun), あたし (atashi), or わたくし (watakushi) with neutral cues. A Bonferroni correction of 4 was applied, causing some $p$ to be greater than 1. ***$p < .001$.}
\label{tab:BreezeGenderComp}
\begin{tabular}{llrrrrr}
\toprule
Assumed distribution & Comparator & $\chi^2$ & Adj. $p$ & Sig. & Comparator $n$ & Effect size $w$ \\ 
\midrule
\multirow[t]{4}{3.25cm}{Neutral: No Self-Referent, わたし (Watashi), and NPSR} & ぼく (Boku) & 5.863 & .474  &   & 204 & 0.170 \\
& じぶん (Jibun) & 5.153 & .644  &   & 204 & 0.159 \\
& あたし (Atashi) & 86.320 & $ < .001$  & ***   & 203 & 0.652 \\
& わたくし (Watakushi)  & 1.354 & 2.865 &  & 204 & 0.082 \\
\bottomrule
\end{tabular}
\end{table*}

For Juniper, we compared the distributions of gender categorisations for the neutral pronouns with the distribution for ぼく (boku) and じぶん (jibun). A Bonferroni correction of 2 was applied, and the expected values were calculated as per \autoref{eq:1}, to account for the different sample sizes.
The results (\autoref{Table:JuniperGenderComp}) show that Juniper using ぼく (boku) was categorised as less feminine and more masculine and ambiguous, \textbf{supporting H5-1}.

\begin{quote}
    \textbf{H5-1:} Using ぼく (boku) with feminine voices will elicit gender-ambiguous impressions of ChatGPT.\\
    $\rightarrow$ \textbf{Confirmed}
\end{quote}

Juniper using じぶん (jibun) elicited similar impressions to the neutral pronouns. Thus:

\begin{quote}
    \textbf{H5-2:} Using じぶん (jibun) with feminine voices will elicit feminine impressions of ChatGPT.\\
    $\rightarrow$ \textbf{Rejected: No differences}
\end{quote}

\begin{table*}[!ht]
\caption{Chi-squared test results comparing the gender perceptions of Juniper when using ぼく (boku) or じぶん (jibun) with neutral cues. A Bonferroni correction of 2 was applied.}
\label{Table:JuniperGenderComp}
\begin{tabular}{llrrrrr}
\toprule
Assumed distribution & Comparator & $\chi^2$ & Adj. $p$ & Sig. & Comparator $n$ & Effect size $w$ \\ \midrule
\multirow[t]{3}{3.25cm}{Neutral: No Self-Referent, わたし (Watashi), and NPSR}              & ぼく (Boku)       & 189.41               & $<.001$ & ***    & 203          & 0.97        \\
 & じぶん (Jibun)      & 3.64               & .606 &      & 204          & 0.13        \\
 & & & & & &  \\
\bottomrule
\end{tabular}
\end{table*}

For Ember, we compared the distributions of gender categorisations for the neutral pronouns to あたし (atashi) and わたくし (watakushi). A Bonferroni correction of 2 was applied, and the expected values were calculated as per \autoref{eq:1}, to account differing sample sizes.
The results (\autoref{Table:EmberGenderComp}) suggest that Ember was considered statistically significantly less masculine and more ambiguous when using あたし (atashi) compared to the neutral pronouns.

\begin{quote}
    \textbf{H6-1:} Using あたし (atashi) with masculine voices will elicit gender-ambiguous impressions of ChatGPT.\\
    $\rightarrow$ \textbf{Confirmed}
\end{quote}

\begin{quote}
    \textbf{H6-2:} Using わたくし (watakushi) with masculine voices will elicit masculine impressions of ChatGPT.\\
    $\rightarrow$ \textbf{Rejected: No differences}
\end{quote}

The open-ended responses provide further insights for RQ2, 
with mentions of sexuality-related terms such as ``gay'' or ``おネエ (onee).'' Specifically, there were 11 for Breeze using あたし (atashi), eight
for Ember using あたし (atashi), and three for Juniper using ぼく (boku) (Table S4 in Supplementary Materials).

\begin{table*}[!ht]
\caption{Chi-squared test results comparing the gender perceptions of Ember when using あたし (atashi) or わたくし (watakushi) with neutral cues. A Bonferroni correction of 2 was applied.}
\label{Table:EmberGenderComp}
\begin{tabular}{llrrrrr}
\toprule
Assumed distribution & Comparator & $\chi^2$ & Adj. $p$ & Sig. & Comparator $n$ & Effect size $w$ \\ \midrule
\multirow[t]{3}{3.25cm}{Neutral: No Self-Referent, わたし (Watashi), and NPSR}               & あたし (Atashi)     & 209.60               & $<.001$ & ***    & 204          & 1.01        \\
 & わたくし (Watakushi)  & 0.42               & 1.872 &     & 204          & 0.05 \\
 & & & & & & \\ \bottomrule
\end{tabular}
\end{table*}

\subsection{Intersectionality and Identity in Japanese Self-Referents (RQ3)}

\begin{figure*}[!ht]
\centering
\includegraphics[width=\textwidth]{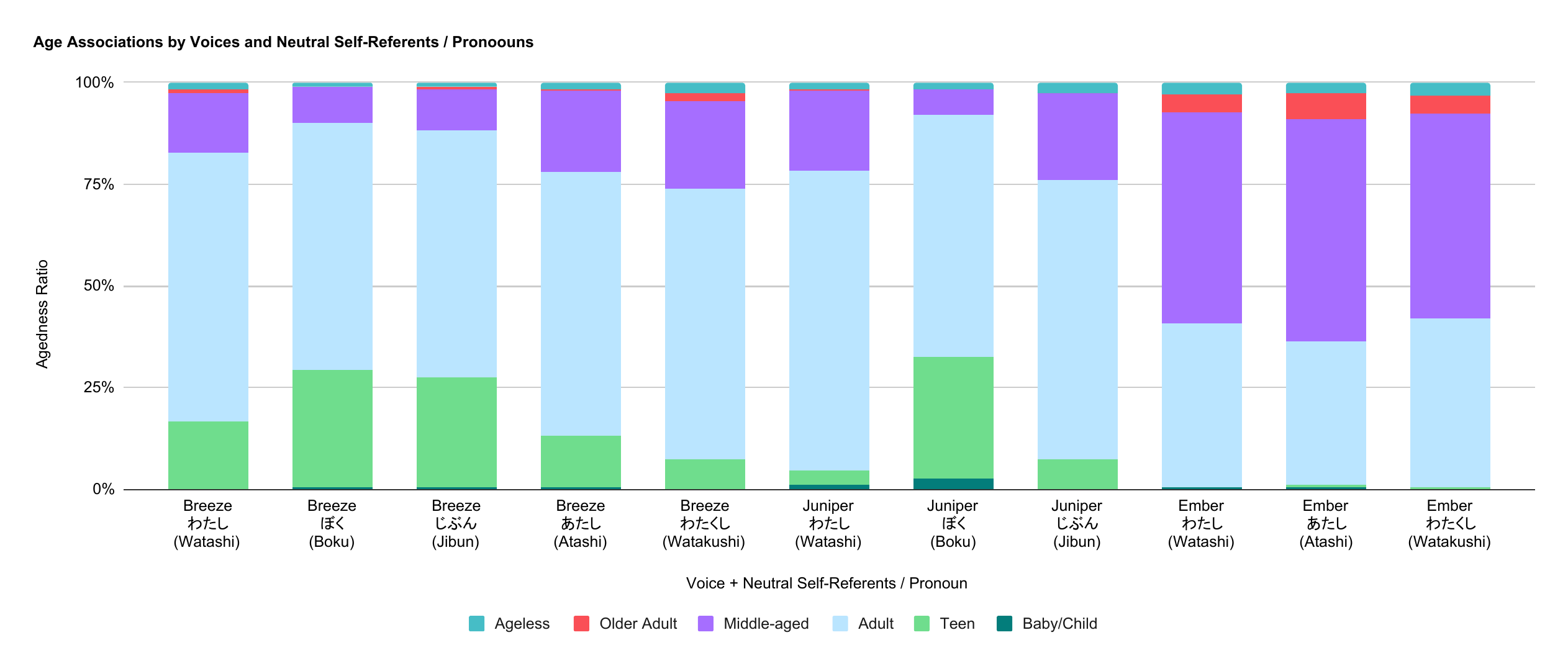}
\caption{Age categorizations for each voice with each self-referent as a percentage ($N=204$).}
\Description{All voices using each pronouns and their assignations of age, including baby/child, teen, adult, middle-aged, older adult, and ageless. Breeze and Juniper had about 60\% associations with adult, and especially Breeze using ぼく (boku) and じぶん (jibun), Juniper using ぼく (boku) had about 30\% associations with teen. Ember with all self-referents had about 50\% associations with middle-aged and 40\% with adult.}
\label{fig:RQ3result_age}
\end{figure*}

For H7, we compared the distributions of age categorisations for NPSR with わたし (watashi) and No self-referent. A Bonferroni correction of 2 was applied.
For Juniper, a statistically significant difference was found, where NPSR was deemed younger than わたし (watashi; $\chi^2(5, 204) = 25.77, p<.001, w=0.36$) and No self-referent ($\chi^2(5, 204) = 17.52, p=.007, w=0.29$). NPSR was generally categorised as a teen or adult, while わたし (watashi) and No self-referent were adult or middle-aged (\autoref{fig:RQ3result_age}).
For Breeze, there were no statistically significant differences between NPSR and わたし (watashi) 
and No self-referent
, with all predominantly categorised as adult.
For Ember, no participants categorised NPSR as baby or child, so no formal analysis was done. The values suggest no difference between NPSR, わたし (watashi), and No self-referent. All three were predominantly categorised as adult (85, 82, and 89 respectively) and middle-aged (103, 106, 99 respectively). Thus:

\begin{quote}
    \textbf{H7:} In the case of NPSR, the age perception of all voices will be younger than when each voice is using わたし (watashi) or No self-referent.\\
    $\rightarrow$ \textbf{Partially supported: Juniper only}
\end{quote}

To test H8, age distributions were compared for each voice by pronoun with わたし (watashi) as the assumed distribution. 
A Bonferroni correction of 2 was applied. The results (\autoref{table:juniperAge}) indicated that ぼく (boku) was deemed statistically significantly younger than わたし (watashi), with Juniper categorised as teen/adult when using ぼく (boku) and adult/middle-aged when using わたし (watashi). No statistically significant difference was found between わたし (watashi) and じぶん (jibun).

\begin{table*}[!ht]
\caption{Chi-squared test results comparing how Juniper was categorised in terms of age when using わたし (watashi) compared to ぼく (boku) and じぶん (jibun). A Bonferroni correction of 2 was applied.}
\label{table:juniperAge}
\begin{tabular}{lllrrrrr}
\toprule
Assumed distribution & Voice   & Comparator & $\chi^2$ & Adj. $p$ & Sig. & Comparator $n$ & Effect size $w$ \\ \midrule
わたし (Watashi)              & Juniper & ぼく (Boku)       & 444.866          & $<.001$ & ***  & 204          & 1.476725        \\
&         & じぶん (Jibun)      & 12.34418               & 0.061 &  & 204          & 0.245989        \\ \bottomrule
\end{tabular}
\end{table*}

For Breeze, there were no categorisations of baby/child for わたし (watashi), preventing testing. The descriptive statistics (\autoref{fig:RQ3result_age}) suggest that わたし (watashi), ぼく (boku), ぶん (jibun), あたし (atashi), and わたくし (watakushi) all had similar teen, adult, and middle-aged categorisations. Similarly, Ember using わたし (watashi) had no teen categorisations, preventing testing. Descriptive statistics (\autoref{fig:RQ3result_age}) suggest that わたし (watashi), あたし (atashi), and わたくし (watakushi) were similarly categorised as adult/middle-aged.
In summary:

\begin{quote}
    \textbf{H8-1:}  The age perception of Breeze and Juniper using ぼく (boku) will be younger than わたし (watashi).\\
    $\rightarrow$ \textbf{Partially supported: Juniper only}
\end{quote}
\begin{quote}
    \textbf{H8-2:}  The age perception of Breeze and Juniper using じぶん (jibun) will be younger than わたし (watashi).\\
    $\rightarrow$ \textbf{Rejected: No differences}
\end{quote}
\begin{quote}
    \textbf{H8-3:}  The age perception of Breeze and Ember using あたし (atashi) will be younger than わたし (watashi).\\
    $\rightarrow$ \textbf{Rejected: \hl{No differences}}
\end{quote}
\begin{quote}
    \textbf{H8-4:}  The age perception of Breeze and Ember using わたくし (watakushi) will be older than わたし (watashi).\\
    $\rightarrow$ \textbf{Rejected: \hl{No differences}}
\end{quote}

For perceptions of Juniper as kawaii, Friedman's tests were run because the data violated assumptions of sphericity. Bonferroni-corrected post-hoc pair-wise comparisons were run.
A statistically significant difference was found among the pronouns ($\chi^2$(4)=$19.17$, $p<.001$). Pairwise comparisons found only one statistically significant difference: between ぼく (boku) and じぶん (jibun) ($p=.036$), where ぼく (boku) was rated as more kawaii ($M=2.96, SD = .838$) than じぶん (jibun) ($M=2.73, SD = .856$).
Thus:
\begin{quote}
    \textbf{H9-1:} Juniper using NPSR will be perceived as more kawaii than Juniper using わたし (watashi), No self-referent, and じぶん (jibun).\\
    $\rightarrow$ \textbf{Rejected: No differences}
\end{quote}

\begin{quote}
    \textbf{H9-2:} Juniper using ぼく (boku) will be perceived as more kawaii than Juniper
using わたし (watashi), No self-referent, and じぶん (jibun).\\
    $\rightarrow$ \textbf{Partially supported: ぼく (boku) was more kawaii than じぶん (jibun)}
\end{quote}

The open response data and code ぼくっこ (boku-kko) was enlightening. For Juniper using ぼく (boku), seven referenced ぼくっこ (boku-kko). No mentions were found for any other pronouns or voices (\autoref{fig:RQ3result_age}). These participants 
gave a mean kawaii score of 3.43, higher than those from other participants (refer to Table S3 in Supplementary Materials), although the sample size differences prevent inferential analysis.

Juniper using じぶん (jibun) received two rural attributions: 
one for ぼく (boku) and four for No self-referent.
Breeze using じぶん (jibun) elicited 12 rural attributions (\autoref{table:thematic}). 
Chi-squared tests indicated statistically significant differences when comparing じぶん (jibun) with No self-referent and わたくし (watakushi) (both two rural attributions; $\chi^2$(1)=$7.40, p=.040, w=.19$- Bonferroni-corrected for six comparisons), but not the other pronouns: seven for ぼく (boku), five for わたし (watashi), eight for NPSR, and three for あたし (atashi). As such:

\begin{quote}
    \textbf{H10-1:} Juniper and Breeze using じぶん (jibun) will elicit a rural image compared to other pronouns.\\
    $\rightarrow$ \textbf{Partially supported: Breeze only, compared to No self-referent and わたくし (watakushi)}
\end{quote}

In terms of formality, Chi-square tests comparing the counts (\autoref{table:thematic}) for Breeze using わたくし (watakushi) with the other self-referents were run, with a Bonferroni correction of 6. The results indicated that わたくし (watakushi) was significantly more formal than: ぼく (boku, $\chi^2(1) = 13.55, p = .001, w = .26$); No self-referent ($\chi^2(4) = 19.51, p < .001, w = .31$); わたし (watashi, $\chi^2(14) = 11.12, p = .001, w = .23$); NPSR  ($\chi^2(1) = 13.55, p = .001, w = .26$), and あたし (atashi, $\chi^2(1) = 14.90, p < .001, w = .27$).
For Ember (37 counts), no statistically significant differences were found: 37 for No self-referent,	23 for わたし (watashi),	35 for NPSR, and 22 for あたし (atashi). Therefore:

\begin{quote}
    \textbf{H10-2:} Ember and Breeze using わたくし (watakushi) will elicit a formal impression compared to each voice using other pronouns.\\
    $\rightarrow$ \textbf{Partially supported: Only for Breeze}
\end{quote}

\begin{table*}[!ht]
\caption{Thematic findings on the personas ($N=204$).}
\label{table:thematic}
\begin{tabular}{p{1.4cm}>{\raggedright\arraybackslash}p{1.4cm}l|r|rr|r>{\raggedleft\arraybackslash}p{1.25cm}>{
\raggedleft\arraybackslash}p{1.25cm}>{\raggedleft\arraybackslash}p{1.25cm}}%
\toprule
&  &  & ぼくっこ & \multicolumn{2}{l} {Region} & \multicolumn{4} {l} {Formality}  \\
& Self-Referent & Voice & \textbf{Boku-kko} & Urban & \textbf{Rural} & \textbf{Formal} & Earnest/ Polite & Smart/ Business & Classy/ Noble \\  \midrule
\multirow [t]{2}{2cm}{First-Person Pronouns} & ぼく (Boku) & Breeze  & 0 & 0 & 7 & 14 & 9 & 5 & 0 \\ 
 & じぶん (Jibun) &  & 0 & 0 & 12 & 17 & 9 & 8 & 0 \\ 
        & あたし (Atashi) & & 0 & 0 & 3 & 13 & 9 & 5 & 0 \\
        & わたくし (Watakushi) & & 0 & 0 & 2 & 39 & 23 & 15 & 1 \\ \cline{2-10} 
 & ぼく (Boku) & Juniper & 7 & 0 & 1 & 15 & 8 & 8 & 0 \\
        & じぶん (Jibun) & & 0 & 0 & 2 & 21 & 6 & 15 & 1 \\  \cline{2-10} 
 & あたし (Atashi) & Ember & 0 & 1 & 0 & 22 & 7 & 16 & 1 \\
        & わたくし (Watakushi) &  & 0 & 0 & 1 & 37 & 18 & 17 & 2 \\ \midrule

\multirow [t]{2}{2cm}{Neutral Self-Referent} & No self-referent & Breeze & 0 & 0 & 2 & 10 & 8 & 2 & 0 \\ 
        & わたし (Watashi) & & 0 & 1 & 5 & 16 & 9 & 8 & 0 \\ 
        & NPSR & & 0 & 0 & 8 & 14 & 9 & 6 & 0 \\ \cline{2-10} 
        & No Self-Referent & Juniper & 0 & 0 & 4 & 21 & 6 & 16 & 0 \\ 
        & わたし (Watashi) & & 0 & 0 & 0 & 19 & 7 & 11 & 1 \\ 
        & NPSR & & 0 & 0 & 0 & 18 & 10 & 8 & 0 \\ \cline{2-10} 
        & No Self-Referent & Ember & 0 & 0 & 0 & 37 & 16 & 21 & 0 \\ 
        & わたし (Watashi) & & 0 & 0 & 0 & 23 & 10 & 15 & 1 \\ 
        & NPSR & & 0 & 0 & 1 & 35 & 11 & 24 & 0 \\ 
 \bottomrule
\end{tabular}
\end{table*}

\subsection{Supplementary Analysis}

112 out of 204 participants (54.9\%) varied in the gender that they attributed to Juniper by self-referent. For Breeze, it was 74 participants (36.3\%), and for Ember, 25 participants (12.3\%). 
This suggests that self-referent did influence perceptions to some extent. In total, 141 participants (69.1\%) varied in how they gendered at least one voice, and 63 (30.9\%) did not: most participants were attending to both the voice and the self-referent. 
Juniper varied the most, suggesting that participants were not biased by the initial self-referent that they experienced. However, we lack a control to confirm. The Juniper results could be a function of voice ambiguity, while Breeze and Ember being more consistently gendered regardless of self-referent could link to the strength of the voice stimulus.

\section{Discussion}
The link between ChatGPT voice and self-referent was nuanced. Gender was a powerful factor, but certain intersectional pronouns and self-referents indicated gender and other forms of identity neutrality, evasiveness, and ambiguity.

\subsection{Gender Identity and Neutrality in Japanese Self-Referents (RQ1)}

Answering this question proved a bit evasive.
Our ``gender-neutral'' voice Breeze was unexpectedly perceived as masculine, despite media attributions implying ambiguity. H1, H2-1, and H3-1 were supported: perceptions of gender-neutral self-referents were influenced by voice gendering. People were primed to gender even when the voice was  ``objectively'' genderless~\cite{Sutton2020ambiguous}. Yet, we might expect a chatbot's ambiguous voice to be gendered feminine due to the ``female by default'' bias~\cite{Jungyong2022stereotypes,unesco2019}.
ChatGPT only provided simple information, which 
may not have evoked the stereotyped associations of femininity with warmth and submissiveness~\cite{unesco2019, curry2020conversational, Jung2023empathic}. Conversely, there may be a masculine default bias, with any agent (human or otherwise) of unknown gender identity assumed ``male'' in the absence of clear gender cues~\cite{bailey2019man}. This may be especially pertinent to the ``Internet'' generation and ChatGPT, which was trained on Internet-based data: research on over 630 billion English words therein indicated that we define ``people'' as ``men''~\cite{bailey2022based}. Ember also overrode the ``neutral'' わたし (watashi), contrasting the text-based results~\cite{fujii2024silver}, perhaps due to its deemed masculine voice. 


Feminine Juniper presented a counterpoint. Juniper was perceived as more feminine when not using a self-referent over the ``feminine when used by women'' わたし (watashi), rejecting H2-2. This suggests that while わたし (watashi) is generally perceived as gender-neutral or feminine~\cite{fujii2024silver, nakamura2014gender, nakamura2022feminist}, the gender associations of a pronoun may be disrupted by a sufficiently feminine voice. 
Here, providing multiple gender cues, which could potentially reinforce biases~\cite{furukawa2013translation}, instead led to a more diverse and nuanced perception of agent gender.
Combining linguistic and voice cues can yield unexpected results, warranting pre-deployment perception tests.

NPSR or illeism was a novel strategy with virtually no prior theory for computer agents, text or voice. We found that the self-referent えーあい (AI) did not further feminize the feminine Juniper voice (H3-2), meaning it was as gender-evasive as having no self-referent. Apparently, the gendered context typically associated with NPSR~\cite{Kojima2017npsr,takahashi2009okinawa, kajino2011,Satoh2021Self, miyazaki2016japanese, Komori2008jishosi} did not apply. However, these results should be confirmed with a variety of feminine agent voices.
Ember was perceived as masculine, which overrode the gender-neutral self-referents (H2-3 and H3-3). This result opposes the ``female by default'' trend
~\cite{unesco2019, Jungyong2022stereotypes} and suggests that the power of voice stimuli can be great, indeed~\cite{McGinn2019voice}. 

\subsection{Gender Evasion by Queering Expectations (RQ2)}
The feminine Juniper using the masculine ぼく (boku) and masculine Ember using the feminine あたし (atashi) successfully queered expectations by eliciting gender-ambiguous perceptions, demonstrating the potential for these pairings to blur conventional gender identities. This aligns with findings on voice gendering exerting great influence on other gender perceptions~\cite{Seaborn2021Voice, Seaborn_2022neutral, Sutton2020ambiguous,torre2023ambig,McGinn2019voice}, and also reveals that the impact of pronoun cues is relative.
For masculine Breeze, the femininity of あたし (atashi)~\cite{nakamura2014gender, miyazaki2016japanese} altered the predominantly masculine voice stimulus, or perhaps broadened the perceived gender spectrum~\cite{Seaborn2022Expansive}. This and the results of H6-1 showing the same phenomenon with Ember confirmed that the influence of あたし (atashi)  found in \cite{fujii2024silver} remains strong.
This queering of self-referents via voice suggests a novel means of amplifying the effects of agent customization through voice variations~\cite{Lotus2021social}.

The findings for feminine Juniper (H5-1) demonstrate that ぼく (boku) remains a strong masculine marker~\cite{ide1990and, nakamura2014gender, fujii2024silver} that either merged with or counteracted voice gendering. This ran counter to the gender-crossing of Breeze and Ember with あたし (atashi). 
Mismatching gender cues might reduce gender stereotypes~\cite{McGinn2019voice, Bernotat2021robot}, so this pairing may be explored in future antistereotyping and debiasing work.
In contrast, じぶん (jibun) had no effect. This syncs with the text-based results, where gender perceptions of じぶん (jibun) were not as strong as ぼく (boku)~\cite{fujii2024silver}, also matching other studies on non-computer interlocutors~\cite{Ogino2007jibun, nakamura2014gender, takahashi2009okinawa}.
H6-2 revealed the same phenomenon for わたくし (watakushi). In contrast to H6-1, わたくし (watakushi) did not overturn the gendering effects of the voice. \citet{fujii2024silver} also showed that the gendered perception of わたくし (watakushi) was not as strong as あたし (atashi), and other work~\cite{nakamura2014gender, nakamura2022feminist} also suggests that わたくし (watakushi) is used in a gender-neutral way, in formal situations. The boundary between ぼく (boku) / あたし (atashi) and じぶん (jibun) / わたくし (watakushi) depends on 
voice gendering, which determines whether じぶん (jibun)/ わたくし (watakushi) can serve as neutral self-referents for voice-based agents.

Using multiple/queer gender markers, while producing gender-evasive results, may also trigger biases related to specific sexualities~\cite{kawano2016onee, maree2013onee, nakamura2022feminist}. Although few, some open-ended responses contained discriminatory content, notably the stereotypical おネエ (onee) image shaped by Japanese media. 
Even simple manipulations of voice and self-referents may (un)expectedly evoke minority personas, warranting careful consideration.

\subsection{Intersectionality and Identity in Japanese Self-Referents (RQ3)}
The elicitation of various social identities was confirmed across the board in ways that intersect with gender. Given that each combination draws out a persona in subtle ways, we now closely examine these nuances in detail.

On age, NPSR was notable when used with Juniper, reflecting the unique image associated with NPSR and little girls~\cite{miyazaki2016japanese, kajino2011, Komori2008jishosi, nakamura2014gender}. While not perceived as kawaii (H9-1), this may be due to a lack of other cues, like appearance.
Still, Juniper using ぼく (boku) \emph{was} perceived as kawaii (H9-2), especially when compared to じぶん (Jibun), reinforcing a ぼくっこ (boku-kko) image~\cite{Nishida2011bokukko}. This combination also received seven mentions of ぼくっこ (boku-kko) by those who evaluated it as kawaii. 
For these people, merely manipulating voice and self-referent crafted a dramatically different persona. This is evidence that simple adjustments can elicit perceptions of diverse, culturally-sensitive personas in computer voice.
In Chinese, the first-person pronoun 人家 (renjia) could be harnessed 
to emphasize girlishness or cuteness~\cite{sandel2018renjia}. This pronoun could function much like the Japanese ぼくっこ (boku-kko), offering translation of our results to other languages.
Similar to previous work~\cite{seaborn_can_2023, seaborn2023kawaiigame}, our results point to intersectional perceptions of kawaii in age and gender through voice. 
Use of ぼく (boku), which, in the text-based work on ChatGPT, was associated with $\sim$50\% teen and $\sim$20\% child impressions~\cite{fujii2024silver}, elicited a teen impression for Juniper and  
a clear ``young boy/girl'' persona~\cite{nakamura2014gender, miyazaki2016japanese}. 
Voice agedness also overrode the effect of pronouns じぶん (jibun) and あたし (atashi), perceived as younger, and わたくし (watakushi), perceived older, compared to わたし (watashi) in the text-based work~\cite{fujii2024silver}. In short, age was the strongest cue. But this was also disruptive: there was no clear single persona~\cite{torre2023ambig}. This may be a form of \emph{social identity elusion} and specifically \emph{gender queering} via age cues.



Rural and regional impressions appeared across various combinations (H10-1). The unclear results for rurality link to the  low counts 
for the Chi-squared test. The agent's use of formal speech in the standard accent may have dampened the pronoun cue. Perhaps the influence of not only pronouns but also the dialect, notably the Kansai dialect~\cite{kori2012dialect}, and intonation~\cite{takagi2005kansai} are needed to enact a regional cue. However, the low association of ruralness for Juniper matches the idea that polite language is expected for women~\cite{kumagai2010dialect,takagi2005kansai, ide1997women}, while using the formal pronoun わたくし (watakushi) aligns with the informality of dialect perceptions~\cite{takagi2005kansai}. These results are subtle and deserve future attention.

On formality, Breeze using わたくし (watakushi) alone elicited a formal impressions, which was theoretically expected~\cite{nakamura2014gender, fujii2024silver}. Additionally, Ember's masculine, deeper voice strongly elicited the impression of a formal persona, even in the absence of わたくし (watakushi). The qualitative data revealed that Breeze consistently elicited impressions of earnestness and politeness across all self-referents and pronouns. In contrast, Ember 
evoked associations with intelligence and capability, particularly in professional or occupational contexts, even when わたくし (watakushi) was not used. 
Masculine agent voices may be more likely to elicit perceptions of intelligence and competence~\cite{bryant2020should}, rather than politeness~\cite{ide1997women, nakamura2022feminist}, indicating a potential gendered bias in the public situations.


\subsection{Implications for Design and Practice}
We offer the following suggestions and guidelines for implementing our findings in voice-based agents and interfaces:

\begin{itemize}
    \item \textbf{Pre-tests and manipulation checks 
    should be conducted to confirm perceptions of social identity factors before proceeding with research or implementation.} Manipulation checks for social identity have recently been highlighted for voice UX work~\cite{Seaborn_2025unbox}. In this study, the self-referents we assumed to be gender-neutral were rendered ineffective by the power of the gendered voices. We should avoid relying on assumptions and be sure to test expected associations.
    
    \item \textbf{Consider that ChatGPT voices evoke distinct personas based on gender and age.} Breeze elicited a teen or adult man persona, Juniper an adult woman persona, and Ember an adult or middle-aged man persona. This knowledge can be harnessed to create more complex and diverse personas or for user personalization.
    
    \item \textbf{Queering combinations can be gender-elusive.} Feminine Juniper using ぼく (boku) (masculine) and masculine Breeze and Ember using あたし (atashi) (feminine) generated ambiguous impressions by incorporating multiple gender markers. 
    However, \textbf{we must be cautious, as queering combinations could also reinforce existing biases,} especially given the associations with how sexual minorities speak and their marginalization.
    \item Juniper using NPSR and ぼく (boku), evoked a youthful impression, while Juniper with わたくし (watakushi) evoked a formal impression, i.e., \textbf{gender, age, and formality}. Impressions of \textbf{kawaii and regional ruralness} were limited to specific combinations of voice and pronouns, namely kawaii for Juniper using ぼく (boku) and regional ruralness for Breeze and Juniper using じぶん (jibun). 
    These elements are ever-interconnected and must be carefully considered when combining voice and self-referent. 
    
\end{itemize}

\subsection{Limitations}
Our most prominent limitation was the Breeze voice failing to be gender neutral or ambiguous, as expected. Practitioners should be cautious about the use of this voice, and perhaps explore あたし (atashi) as a mediator. 
The unclear results for region may be due to not controlling for dialectical features beyond self-referents and/or participant groups, which the text-based study did~\cite{fujii2024silver}. Synthetic speech accents can trigger negative stereotypes associated with the regional impression of a given agent~\cite{Lotus2021social}. Thus, we should consider prosodic speech variations beforehand.

The supplementary analysis indicated that participants tended not to form a single identity perception per voice based on the first voice and self-referent combination experienced. Yet, the results for Ember and Breeze indicated there was less variation. This suggests that randomizing the presentation order was somewhat insufficient. A between-subjects design can verify the results. Still, such a study may be affected by individual differences; an alternative could track which voice-pronoun pair was presented first for each participant. 

An interactive (non-video) study will be needed. People are becoming increasingly accustomed to communicating with voice agents in their daily lives~\cite{Lotus2021social}. For example, (non-)verbal cues like nodding and agreement can influence emotions and behaviour in conversations with virtual humans~\cite{gratch2013using}. Future replications may also find enhanced or broader perceptions of specific personas in an interactive setting. 
Also, the elusive personas found in this study could be employed in virtual characters to mitigate discriminatory perceptions and prejudices, as in \citet{haake2008visual}.

Future research should explore other languages and linguistic phenomena. In Japanese, sentence-ending words also vary according to gender, age, and social class~\cite{nakamura2014gender,nakamura2022feminist}. First-person pronouns and gender binary sentence-ending words also vary in the Thai language~\cite{Saisuwan2015Thai} and Chinese via 人家 (renjia)~\cite{sandel2018renjia}. 
The impact of language-specific speech cues used by voice agents needs urgent, culturally aware study.

\section{Conclusion}
Japanese self-referents and pronouns are important identity markers and their integration with voice has generated novel insights to design LLM-based agents. By carefully combining voice with self-referents, we successfully elicited gender-evasive perceptions in ChatGPT. This approach reveals intersections in gender, age, kawaii, region, and formality. 
We offer a simple and effective way to evoke diverse personas through voice in LLM-based agents. Exploring a gender-neutral alternative to Breeze and nuanced speech elements, such as accent, are for future work to explore.


\end{CJK}

\begin{acks}
This work was funded by a Japan Society for the Promotion of Science (JSPS) Grants-in-Aid for Scientific Research B (KAKENHI Kiban B) grant (\#24K02972). We thank Suzuka Yoshida and the Aspirational Computing Lab for pilot testing and research assistance.
The authors used Deepl and ChatGPT to translate, shorten, and check select passages to and from Japanese and English.
\end{acks}

\balance
\bibliographystyle{ACM-Reference-Format}
\begin{CJK}{UTF8}{ipxm}
\bibliography{refs}
\end{CJK}

\end{document}